\documentclass[prd,preprint,amsmath,amssymb,superscriptaddress,floatfix,nofootinbib,11pt]{revtex4-1}
\usepackage[utf8]{inputenc}
\usepackage{mathrsfs,bm}
\usepackage[sort&compress]{natbib}
\usepackage{txfonts}
\usepackage{amssymb}
\usepackage{rotating}
\usepackage{indentfirst}
\usepackage{graphicx,booktabs}
\usepackage{multirow}
\usepackage{slashed}
\usepackage{overpic}
\usepackage{color}
\usepackage{amssymb}
\usepackage{hyperref}
\usepackage{bbding}
\usepackage{url}

\begin{document}

\title{$V_{cb}$ puzzle in semi-leptonic $B\to D^*$ decays revisited}

\author{Shuang-Yi Li}
\affiliation{School of Physics, Beihang University, Beijing 102206, China}

\author{Jie Xu}
\affiliation{School of Physics, Beihang University, Beijing 102206, China}

\author{Rui-Xiang Shi}
\email[Corresponding author: ]{ruixiang.shi@gxnu.edu.cn}
\affiliation{Department of Physics, Guangxi Normal University, Guilin 541004, China}
\affiliation{Guangxi Key Laboratory of Nuclear Physics and Technology, Guangxi Normal University, Guilin 541004, China}

\author{Li-Sheng Geng}
\email[Corresponding author: ]{lisheng.geng@buaa.edu.cn}
\affiliation{School of Physics, Beihang University, Beijing 102206, China}
\affiliation{Sino-French Carbon Neutrality Research Center, \'Ecole Centrale de P\'ekin/School of General Engineering, Beihang University, Beijing 100191, China}
\affiliation{Peng Huanwu Collaborative Center for Research and Education, Beihang University, Beijing 100191, China}
\affiliation{Beijing Key Laboratory of Advanced Nuclear Materials and Physics, Beihang University, Beijing 102206, China }
\affiliation{Southern Center for Nuclear-Science Theory (SCNT), Institute of Modern Physics, Chinese Academy of Sciences, Huizhou 516000, China}

\author{Yi Zhang}
\affiliation{School of Physics, Beihang University, Beijing 102206, China}

\begin{abstract}
Inspired by the newly reported $B\to D^*(\to D\pi)\ell\bar{\nu}_\ell$ differential decay rates by the Belle and Belle II Collaborations,  we revisit the $V_{cb}$ puzzle in semi-leptonic $B\to D^*$ decays, considering the latest lattice QCD~(LQCD) simulations and light-cone sum rule~(LCSR) results. We examine the commonly used Caprini-Lellouch-Neubert~(CLN), Boyd-Grinstein-Lebed~(BGL), and heavy quark effective theory~(HQET) parameterizations. We demonstrate that these three parameterizations yield consistent results and reconfirm the $V_{cb}$ puzzle. Then, we use a state-of-the-art Bayesian method to estimate the impact of higher-order terms beyond the present HQET expansion on the uncertainty of $V_{cb}$. 
We show that higher-order effects cannot eliminate the deviation between the exclusive and inclusive determinations of $V_{cb}$. Finally, utilizing the best-fit results obtained in the HQET parameterization via fitting LQCD and LCSR data only as inputs, we predict the relevant observables, i.e., $R_{D^*}$, $F_L^{D^*}$, and $P_\tau^{D^*}$, sensitive to new physics in the $B\to D^*\ell\bar{\nu}_\ell$ decays. We conclude that lepton-flavour universality violations still exist in the $b\to c\tau\nu$ transitions. 
\end{abstract}


\maketitle


\section{Introduction}
The precise determination of the Cabibbo–Kobayashi–Maskawa~(CKM) matrix elements plays an important role in testing the Standard Model~(SM) and searching for new physics~(NP) beyond the SM~(BSM). The $|V_{cb}|$ determination in semi-leptonic $B\to D^*$ decays has attracted great attention in the past decade due to the long-standing $V_{cb}$ puzzle~\cite{HFLAV:2022esi,FlavourLatticeAveragingGroupFLAG:2024oxs,Bernlochner:2024sfg}, i.e., the $|V_{cb}|$ value determined from exclusive decays deviates from that from inclusive decays. 

To investigate the issue, one must understand the experimental differential distributions and form factors associated with the $B\to D^*$ transition. In 2017, the Belle Collaboration first measured the differential decay distribution of $B\to D^*\ell\bar{\nu}_\ell$ using a hadronic tag~\cite{Belle:2017rcc}. Subsequently, combined with the LQCD form factors at zero recoil~\cite{FermilabLattice:2014ysv}, LCSR form factors at maximal recoil~\cite{Faller:2008tr}, and QCD sum rule~(QCDSR) calculations~\cite{Neubert:1992wq,Neubert:1992pn,Ligeti:1993hw}, several groups~\cite{Grinstein:2017nlq,Bigi:2017njr,Bernlochner:2017jka,Jaiswal:2017rve,Bigi:2017jbd,Jung:2018lfu,Bernlochner:2017xyx,Bernlochner:2019ldg} studied the $|V_{cb}|$ determinations in exclusive semi-leptonic $B\to D^*$ decays using the CLN~\cite{Caprini:1997mu}, BGL~\cite{Boyd:1997kz}, and next-to-leading order~(NLO) HQET~\cite{Neubert:1993mb} parameterizations for the form factors. Surprisingly, the results are inconsistent, and only the $|V_{cb}|$ value determined with the BGL parameterization can reconcile the $V_{cb}$ puzzle. 

In 2018, the Belle Collaboration published the $q^2$ and angular distributions of $B\to D^*\ell\bar{\nu}_\ell$ obtained from untagged analyses~\cite{Belle:2018ezy}. That same year, Gubernari et al.~\cite{Gubernari:2018wyi} improved the results of Ref.~\cite{Faller:2008tr} by calculating the $B\to D^*$ form factors up to twist four distribution amplitudes using LCSR. These allowed updated analyses of the $|V_{cb}|$ extraction and its parametrization dependence. Unlike the conclusion reached in Ref.~\cite{Grinstein:2017nlq,Bigi:2017njr} based on the 2017 dataset, the updated works with the CLN and BGL parameterizations~\cite{Gambino:2019sif,Jaiswal:2020wer,Ferlewicz:2020lxm} showed no sign of parametrization dependence in $|V_{cb}|$ using the 2018 dataset~\cite{Belle:2018ezy}, which still does not solve the $V_{cb}$ puzzle. Other works~\cite{Bordone:2019vic,Iguro:2020cpg,Bernlochner:2022ywh} employed a refined HQET parameterization, considering ${\cal O}(1/m_c^2)$ corrections to the heavy quark expansion, and reached a similar conclusion. To further investigate the dependence of the $|V_{cb}|$ value on different parameterizations and the $V_{cb}$ puzzle, more experimental data and accurate form factors from LQCD and LCSR are needed.

Recently, the Belle Collaboration reported a new measurement of the differential distributions of exclusive $B\to D^*\ell\bar{\nu}_\ell$ decays with a hadronic tag~\cite{Belle:2023bwv}, which supersedes the 2017 Belle result~\cite{Belle:2017rcc}. A similar analysis is also presented based on the Belle II data~\cite{Belle-II:2023okj}. In addition, the LQCD calculations of form factors at nonzero recoil~\cite{FermilabLattice:2021cdg,Harrison:2023dzh,Aoki:2023qpa} and improved LCSR results~\cite{Cui:2023jiw,Gubernari:2018wyi} on the $B\to D^*\ell\bar{\nu}_\ell$ form factors have become available.  Several studies~\cite{Bordone:2024weh,Martinelli:2023fwm,Kapoor:2024ufg} have employed a subset of the aforementioned data to perform a re-examination of $|V_{cb}|$ value using the BGL approach or the Dispersive Matrix method~\cite{Lellouch:1995yv,DiCarlo:2021dzg} for the form factors. Although these studies still fall short of solving the $V_{cb}$ puzzle, they have demonstrated that different types of data can exert varying degrees of influence on the central value and uncertainty of $|V_{cb}|$. In the bargain, it remains unexplored whether the  $|V_{cb}|$ determination, based on all the most up-to-date data, is contingent upon different parameterization methods.

On the other hand, studies~\cite{Bordone:2019vic,Iguro:2020cpg,Bernlochner:2022ywh} on the  $|V_{cb}|$ determination within the framework of HQET approach have shown that incorporating high-order contributions can alleviate the $V_{cb}$ puzzle. Although the HQET, as an effective field theory~(EFT), holds the promise of systematic expansions, its practical application is hindered by the inability to constrain the increasing number of free parameters at higher orders with the limited data. Typically, the uncalculated higher-order contributions are treated as truncation errors, which dominate the uncertainties in EFT calculations. The rigorous handling of theoretical uncertainties is crucial for exploring high-energy phenomenology~\cite{Cacciari:2011ze}. However, in previous HQET studies\cite{Neubert:1993mb,Bordone:2019vic,Iguro:2020cpg,Bernlochner:2022ywh}, the theoretical uncertainty of $|V_{cb}|$ is generally left unquantified and lacks a well-defined statistical interpretation. The traditional method estimates theoretical uncertainties in EFT calculations through renormalization scale variation and its core defect is the inability to quantify the credibility of uncertainty intervals~\cite{Epelbaum:2019kcf}. In recent years, the state-of-the-art Bayesian framework proposed in Refs~\cite{Furnstahl:2015rha,Melendez:2017phj,Melendez:2019izc} provides a rigorous probabilistic interpretation, addressing the lack of statistical rigor in traditional methods, which has been widely used to study truncation uncertainties in EFTs~\cite{ Lu:2021gsb,Zhai:2022ied,Lu:2022hwm}. 

In the present work, we use the latest Belle and Belle II data and form factors from LQCD and LCSR for $B\to D^*\ell\bar{\nu}_\ell$ as inputs to revisit the $V_{cb}$ puzzle and examine the parameterization dependence on the  $|V_{cb}|$ extraction. In particular, some analyses are performed for investigating the influence of various available data on the $|V_{cb}|$ determination. In addition, for the first time, we study the truncation uncertainties of the HQET expansion on the $V_{cb}$ extraction using the Beysian method~\cite{Furnstahl:2015rha,Melendez:2017phj,Melendez:2019izc}. We show
that incorporating higher-order effects into the HQET cannot solve the $V_{cb}$ puzzle.

This work is organized as follows. In Sec. II, we provide
 the theoretical ingredients for the $|V_{cb}|$ determination in semi-leptonic $B\to D^*$ decays. Results and discussions are given in Sec. III, followed by a summary in Sec. IV.

\section{Theoretical framework}

This section briefly reviews the essential theoretical ingredients to study the $V_{cb}$ puzzle. These include the differential decay distribution for the four-body $B\to D^*(\to D\pi)\ell\bar{\nu}_{\ell}$, the three parameterizations of hadronic form factors on the $B\to D^*$ transition, and the Bayesian method used to estimate higher-order contributions neglected in the HQET.

\subsection{Differential decay distribution for the four-body $B\to D^*(\to D\pi)\ell\bar{\nu}_{\ell}$}

Concerning the available experimental data for the $|V_{cb}|$ determination, one should know the differential decay distribution for the four-body $B\to D^*(\to D\pi)\ell\bar{\nu}_{\ell}$ in the SM, which reads~\cite{Bhattacharya:2022bdk}
\begin{eqnarray}
\frac{d^4 \Gamma}{d q^2 d \cos \theta^* d \cos \theta_{\ell} d \chi}&=&\frac{9}{32 \pi}\left[\left(I_1^s \sin ^2 \theta^*+I_1^c \cos ^2 \theta^*\right)+\left(I_2^s \sin ^2 \theta^*+I_2^c \cos ^2 \theta^*\right) \cos 2 \theta_{\ell}\right.\nonumber\\
&&+I_3 \sin ^2 \theta  ^* \sin ^2 \theta_{\ell} \cos 2 \chi+I_4 \sin 2 \theta^* \sin 2 \theta_{\ell} \cos \chi+I_5 \sin 2 \theta^* \sin \theta_{\ell} \cos \chi
\nonumber\\
&&+\left(I_6^c \cos ^2 \theta^*+I_6^s \sin ^2 \theta^*\right) \cos \theta_{\ell}+I_7 \sin ^2 \theta^* \sin \theta_{\ell} \sin \chi \nonumber\\
&&\left.+I_8 \sin 2 \theta^* \sin 2 \theta_{\ell} \sin  \chi+I_9 \sin ^2 \theta^* \sin ^2 \theta_{\ell} \sin 2 \chi\right],
\end{eqnarray}
with
\begin{equation}
\begin{array}{ll}
I_1^c={\cal N}\left\{4\frac{m_{\ell}^2}{q^2}\left|{H}_t\right|^2+2\left(1+\frac{m_{\ell}^2}{q^2}\right)\left|{H}_0\right|^2\right\}, &
I_1^s=\frac{3}{2}{\cal N}\left(\left|{H}_{+}\right|^2+\left|{H}_{-}\right|^2\right), \\
I_2^c=-2{\cal N}\left(1-\frac{m_{\ell}^2}{q^2}\right)\left|{H}_0\right|^2,  &
I_2^s=\frac{1}{2}{\cal N}\left(1-\frac{m_{\ell}^2}{q^2}\right)\left(\left|{H}_{+}\right|^2+\left|{H}_{-}\right|^2\right), \\
I_3=-2{\cal N}\left(1-\frac{m_t^2}{q^2}\right)H_+ H_-, &
I_4=-{\cal N}\left(1-\frac{m_{\ell}^2}{q^2}\right){H}_0 (H_++H_-), \\ 
I_5= 2{\cal N}\left[{H}_0 (H_+-H_-)-\frac{m_{\ell}^2}{q^2}(H_++H_-) {H}_t\right],&
I_6^c=-8{\cal N}\frac{m_{\ell}^2}{q^2} {H}_0 {H}_t, \\ 
I_6^s=-2{\cal N}(H_+^2-H_-^2), &
I_7=I_8=I_9=0, \\ 
{\cal N}=\frac{G_F^2\eta_{ew}\left|V_{c b}\right|^2\left(q^2-m_{\ell}^2\right)^2\left|\vec{p}_{D^*}\right|}{192 \pi^3 m_B^2 q^2} \mathcal{B}\left(D^* \rightarrow D \pi\right).\label{CoffIi}
\end{array}
\end{equation}
where $\theta_{\ell}$ is the angle  between the opposite direction of the $D^*$ meson and the $\ell^-$ lepton in the dilepton rest frame, $\theta^*$ is the angle between the $D^*$ momentum and its daughter $D$ meson, $\chi$ is the angle between the plane ($D\pi$) and dilepton rest frames, and $q^2$ is the invariant mass of the lepton-neutrino pair. $\cal{N}$ is the normalization factor, $G_F$ is the Fermi constant, $V_{cb}$ is the Cabibbo-Kobayashi-Maskawa (CKM) matrix
element, $\eta_{ew}=1.0066$  represents the leading-order electroweak correction to the SM contribution~\cite{Sirlin:1981ie}, $I_{1-9}^{(s,c)}(q^2)$ are angular coefficients, $\left|\vec{p}_{D^*}\right|=\sqrt{\lambda\left(m_B^2, m_{D^*}^2, q^2\right)} /\left(2 m_B\right)$ is the 3-momentum of the $D^*$ meson, and $\lambda(a, b, c)=$ $a^2+b^2+c^2-2 a b-2 b c-2 c a$. The helicity amplitudes $H_{0,t,\pm}$ are functions of four form factors, $V(q^2)$, $A_0(q^2)$, $A_1(q^2)$, and $A_2(q^2)$ relevant to the $B\to D^*$ transition, as follows:
\begin{eqnarray}
{H}_0&=&-\frac{\left(m_B+m_{D^*}\right)}{2 m_{D^*} \sqrt{q^2}}\left[\left(m_B^2-m_{D^*}^2-q^2\right) A_1\left(q^2\right)-\frac{\lambda\left(m_B^2, m_{D^*}^2, q^2\right)}{\left(m_B+m_{D^*}\right)^2} A_2\left(q^2\right)\right] \nonumber, \\
{H}_t&=&- \frac{\sqrt{\lambda\left(m_B^2, m_{D^*}^2, q^2\right)}}{\sqrt{q^2}} A_0\left(q^2\right), \nonumber \\
{H}_{ \pm}&=&\left(m_B+m_{D^*}\right) A_1\left(q^2\right) \mp\frac{\sqrt{\lambda\left(m_B^2, m_{D^*}^2, q^2\right)}}{m_B+m_{D^*}} V\left(q^2\right).\label{HeliAmp} 
\end{eqnarray}
From Eqs.~(\ref{CoffIi}) and (\ref{HeliAmp}), one can see that the $H_t$ is determined solely by the form factor $A_0(q^2)$ and is multiplied by the lepton mass $m_\ell$. As a result, the contributions of $H_t$ and $A_0(q^2)$ to the $B\to D^*(\to D\pi)\ell\bar{\nu}_{\ell}$ decays will vanish once the lepton mass is neglected. In this work, we refrain from discussing new physics.

\subsection{Hadronic form factor parameterizations}
To extract the $|V_{cb}|$ value in the $B\to D^*(\to D\pi)\ell\bar{\nu}_{\ell}$ decays, one needs to know the form factors in Eq.~(\ref{HeliAmp}). The theoretical descriptions of these form factors are commonly performed in three different parameterizations, namely the CLN~\cite{Caprini:1997mu}, BGL~\cite{Boyd:1997kz}, and HQET~\cite{Neubert:1993mb} parameterizations. 

The CLN parameterization considers dispersion relations and relations at $1/m_b$ in the heavy
quark expansion and uses only four parameters to describe the differential decay distribution. From this perspective, the CLN parameterization can be regarded as a simplified parametrization neglecting higher-order contributions in the heavy
quark expansion. Following the CLN parametrization~\cite{Caprini:1997mu}, one can re-write the form factors in Eq.~(\ref{HeliAmp}) as:
\begin{eqnarray}
&&V(\omega)=\frac{m_B+m_{D^*}}{2\sqrt{m_Bm_{D^*}}}R_1(\omega)h_{A_1}(\omega),\nonumber\\
&&A_0(\omega)=\frac{m_B+m_{D^*}}{2\sqrt{m_Bm_{D^*}}}R_0(\omega)h_{A_1}(\omega),\nonumber\\
&&A_1(\omega)=\frac{\sqrt{m_Bm_{D^*}}}{m_B+m_{D^*}}(\omega+1)h_{A_1}(\omega), \nonumber\\
&&A_2(\omega)=\frac{m_B+m_{D^*}}{2\sqrt{m_Bm_{D^*}}}R_2(\omega)h_{A_1}(\omega),
\end{eqnarray}
with
\begin{eqnarray}
&&h_{A_1}(\omega)=h_{A_1}(1)\left[1-8\rho_{D^*}^2z+(53\rho_{D^*}^2-15)z^2-(231\rho_{D^*}^2-91)z^3\right],\nonumber\\
&&R_1(\omega)=R_1(1)-0.12(\omega-1)+0.05(\omega-1)^2,\nonumber\\
&&R_2(\omega)=R_2(1)+0.11(\omega-1)-0.06(\omega-1)^2,\nonumber\\
&&R_0(\omega)=R_0(1)-0.11(\omega-1)+0.01(\omega-1)^2.
\end{eqnarray}

where $\omega=\frac{m_B^2+m_{D^*}^2-q^2}{2m_Bm_{D^*}}$, and $z=\frac{\sqrt{\omega+1}-\sqrt{2}}{\sqrt{\omega+1}+\sqrt{2}}$. The four unknown parameters $\rho_{D^*}^2$, $h_{A_1}(1)$, $R_1(1)$, and $R_2(1)$ are determined by fitting to the experimental data and LQCD simulations, and $R_0(1)$ can be calculated using HQET~\cite{Fajfer:2012vx}.

The BGL parameterization relies on dispersion relations and operator product expansion, which is a model-independent parameterization. However, the BGL parameterization includes a larger number of unknown parameters which need to be determined by experimental data. Under the BGL parameterization~\cite{Boyd:1997kz}, the helicity amplitudes $H_t$, $H_0$, and $H_\pm$ in Eq.~(\ref{HeliAmp}) can be expressed in terms of redefined form factors $f$, $g$, $\mathcal{F}_1$, and $\mathcal{F}_2$,
\begin{eqnarray}
&&H_t(\omega)=\mathcal{F}_2(\omega),\nonumber\\
&&H_0(\omega)=\mathcal{F}_1(\omega)/\sqrt{q^2},\nonumber\\
&&H_{\pm}(\omega)=f(\omega)\mp m_Bm_{D^*}\sqrt{\omega^2-1}g(\omega),
\end{eqnarray}
and the form factors $f$, $g$, $\mathcal{F}_1$, and $\mathcal{F}_2$ can be further expanded via $z$, which is 
\begin{eqnarray}
F_i(z)&=&\frac{1}{P_{F_i}(z)\phi_{F_i}(z)}\sum_{j=0}^Na_j^{F_i}z^j,
\end{eqnarray}
where $F_i\equiv\{f,g,\mathcal{F}_1,\mathcal{F}_2\}$ and 
$z=\frac{\sqrt{\omega+1}-\sqrt{2}}{\sqrt{\omega+1}+\sqrt{2}}$. $P_{F_i}(z)$ are the Blaschke factors, which read
\begin{eqnarray}
P_{F_i}(z)&=&\prod_{P=1}^n\frac{z-z_P}{1-zz_P},
\end{eqnarray}
with
\begin{eqnarray}
z_P&=&\frac{\sqrt{t_+-m_P^2}-\sqrt{t_+-t_-}}{\sqrt{t_+-m_P^2}+\sqrt{t_+-t_-}},\qquad t_{\pm}=(m_B\pm m_{D^*})^2.
\end{eqnarray}
They can help eliminate poles for $q^2<(m_B+m_{D^*})^2$ associated with the on-shell production of the $B_c^*$ bound states.
The $\phi_i(z)$ are so-called outer functions, defined as
\begin{eqnarray}
&&\phi_f(z)=\frac{4r_{D^*}}{m_B^2}\sqrt{\frac{n_I}{3\pi\chi_{1^+}^T(0)}}\frac{(1+z)(1-z)^{3/2}}{[(1+r_{D^*})(1-z)+2\sqrt{r_{D^*}}(1+z)]^4},\nonumber\\
&&\phi_g(z)=16r_{D^*}^2\sqrt{\frac{n_I}{3\pi\tilde{\chi}_{1^-}^T(0)}}\frac{(1+z)^2(1-z)^{-1/2}}{[(1+r_{D^*})(1-z)+2\sqrt{r_{D^*}}(1+z)]^4},\nonumber\\
&&\phi_{\mathcal{F}_1}(z)=\frac{4r_{D^*}}{m_B^3}\sqrt{\frac{n_I}{6\pi\chi_{1^+}^T(0)}}\frac{(1+z)(1-z)^{5/2}}{[(1+r_{D^*})(1-z)+2\sqrt{r_{D^*}}(1+z)]^5},\nonumber\\
&&\phi_{\mathcal{F}_2}=8 \sqrt{2} r_{D^*}^2 \sqrt{\frac{n_I}{\pi \tilde{\chi}_{1^{+}}^L(0)}} \frac{(1+z)^2(1-z)^{-\frac{1}{2}}}{[(1+r_{D^*})(1-z)+2 \sqrt{r_{D^*}}(1+z)]^4},
\end{eqnarray}
where $r_{D^*}=m_{D^*}/m_B$, relevant inputs  $\chi_{1^+}^T(0), \tilde{\chi}_{1^-}^T(0)$, $\tilde{\chi}_{1^{+}}^L(0)$ and $n_I$ for the outer functions are given in the Appendix~\ref{App:Inputs}. The coefficients $a_j^{F_i}$ satisfy the following weak unitarity constraints
\begin{eqnarray}
\sum_{j=0}^N(a_j^g)^2<1,\qquad
\sum_{j=0}^N(a_j^f)^2+(a_j^{\mathcal{F}_1})^2<1,\qquad
\sum_{j=0}^N\left(a_j^{\mathcal{F}_2}\right)^2<1.
\end{eqnarray}

Note that form factors in Eq.~(\ref{HeliAmp}) are not completely independent. They satisfy the condition, $A_0(\omega_{\rm max})=\frac{m_B+m_{D^*}}{2m_{D^*}}A_1(\omega_{\rm max})-\frac{m_B-m_{D^*}}{2m_{D^*}}A_2(\omega_{\rm max})$ , to cancel the divergence at the maximum recoil ($\omega=\omega_{\rm max}$) in the definition of axial-vector operator matrix element~\cite{Sakaki:2013bfa}. This leads to the following kinematical constraint on the BGL form factors $F_i$, which is 
\begin{eqnarray}
\mathcal{F}_2\left(\omega_{\max }\right)=\frac{1+r_{D^*}}{m_B^2\left(1+\omega_{\max }\right)\left(1-r_{D^*}\right) r_{D^*}} \mathcal{F}_1\left(\omega_{\max }\right).
\end{eqnarray}
In addition, due to the redefinition of form factors in Eq.~(\ref{HeliAmp}) under BGL parameterization, there is an additional kinematic constraint between form factors $F_i$ at $\omega=1$: 
\begin{eqnarray}
\mathcal{F}_1(1)=(m_B-m_{D^*})f(1).
\end{eqnarray}

The HQET parameterization is fully based on heavy quark symmetry. As a result, it allows us to quantitatively study the impact of the contribution of heavy quark expansion at each order on the $V_{cb}$ extraction and to perform a combined analysis of $B\to D$ and $B\to D^*$ processes. In the HQET basis, the form factors in Eq.~(\ref{HeliAmp}) are defined as
\begin{eqnarray}
V\left(\omega\right)&=&\frac{m_B+m_{D^*}}{2 \sqrt{m_B m_{D^*}}} h_V\left(\omega\right), \nonumber\\
A_1\left(\omega\right)&=&\frac{\left(m_B+m_{D^*}\right)^2-q^2}{2 \sqrt{m_B m_{D^*}}\left(m_B+m_{D^*}\right)} h_{A_1}\left(\omega\right), \nonumber\\
A_2\left(\omega\right)&=&\frac{m_B+m_{D^*}}{2 \sqrt{m_B m_{D^*}}}\left[h_{A_3}\left(\omega\right)+\frac{m_{D^*}}{m_B} h_{A_2}\left(\omega\right)\right],\nonumber\\
A_0\left(\omega\right)&=&\frac{1}{2 \sqrt{m_B m_{D^*}}}\left[\frac{\left(m_B+m_{D^*}\right)^2-q^2}{2 m_{D^*}} h_{A_1}\left(\omega\right)\right.\nonumber \\
&&\left.-\frac{m_B^2-m_{D^*}^2+q^2}{2 m_B} h_{A_2}\left(\omega\right)-\frac{m_B^2-m_{D^*}^2-q^2}{2 m_{D^*}} h_{A_3}\left(\omega\right)\right].\label{Eq:HQETFFs}
\end{eqnarray}
In Eq.~(\ref{Eq:HQETFFs}), $h_X(\omega)$ with $X=V,A_1,A_2,A_3$ represent the form factors in the HQET and can be  expressed in terms of the leading Isgur-Wise function $\xi(\omega)$~\cite{Isgur:1989vq} and its correction $\hat{h}_X(\omega)$, which reads
\begin{eqnarray}
h_X(\omega)=\xi(\omega)\hat{h}_X(\omega),
\end{eqnarray}
with
\begin{eqnarray}
\hat{h}_X(\omega)&=&\hat{h}_{X,0}+\epsilon_a\delta\hat{h}_{X,\alpha_s}+\epsilon_b\delta\hat{h}_{X,m_b}+\epsilon_c\delta\hat{h}_{X,m_c}+\epsilon_c^2\delta\hat{h}_{X,m_c^2},
\end{eqnarray}
where $\hat{h}_{X, 0}=1$ for the $X=V,A_1,A_3$, or else $\hat{h}_{X, 0}=0$. Three quantities $\delta\hat{h}_{X,\alpha_s}$, $\delta\hat{h}_{X,m_b}$, $\delta\hat{h}_{X,m_c}$ stand for the NLO corrections in the $\alpha_s$ and $1/m_{b,c}$ expansions. Therein, $\delta\hat{h}_{X,m_b}$ and $\delta\hat{h}_{X,m_c}$ can be expressed in terms of three unknown NLO Isgur-Wise functions $\eta(\omega)$, $\hat{\chi}_2(\omega)$, and $\hat{\chi}_3(\omega)$. $\delta\hat{h}_{X,m_c^2}$ indicates the next-to-next-to-leading-order~(NNLO) corrections in the $1/m_{c}^2$ expansions. At the same time, there are six NNLO Isgur-Wise functions $\hat{\ell}_{1-6}(\omega)$ for $\delta\hat{h}_{X,m_c^2}$. The complete expressions for corrections beyond the leading order are collected in Appendix~\ref{App:HQETcorr}. In this work, we take the expansion coefficients as $\epsilon_a=\alpha_s/\pi=0.0716$, $\epsilon_b=\bar{\Lambda}/(2m_b)=0.0522$, and $\epsilon_c=\bar{\Lambda}/(2m_c)=0.1807$~\cite{Iguro:2020cpg}. 

Following Refs.~\cite{Bordone:2019vic,Iguro:2020cpg}, the Isgur-Wise functions at each order can be expanded in terms of $z$ around $z=0$, namely,
\begin{eqnarray}
f(\omega)=f^{(0)}+8 f^{(1)} z+16\left(f^{(1)}+2 f^{(2)}\right) z^2+\frac{8}{3}\left(9 f^{(1)}+48 f^{(2)}+32 f^{(3)}\right) z^3+{\cal{O}}(z^4),
\end{eqnarray}
where $z=\frac{\sqrt{\omega+1}-\sqrt{2}}{\sqrt{\omega+1}+\sqrt{2}}$, $f=\{\xi$, $\eta$, $\hat{\chi}_2$, $\hat{\chi}_3$, $\hat{\ell}_i$\}. The expansion coefficients of the Isgur-Wise functions can be determined by fitting the available data. In this work, the contributions of the form factors in the HQET parameterization are considered up to ${\cal O}(1/m_c^2)$, referred to as NNLO. As pointed out in Ref.~\cite{Bernlochner:2022ywh}, other ${\cal{O}}(1/m_bm_c,1/m_b^2,\alpha_s/m_{b,c})$ corrections at NNLO are not sensitive to the available data and can be safely neglected for the $|V_{cb}|$ extraction in HQET. To determine the unknown expansion coefficients, we choose the so-called HQET~(2/1/0) model~\cite{Bordone:2019vic,Iguro:2020cpg}. In Refs.~\cite{Bordone:2019vic,Iguro:2020cpg}, the HQET~(2/1/0) has been proven to be the minimal fit model that can achieve a good description of existing data. Here, 2,1 and 0 denote the expansion order of leading, sub-leading, and sub-sub-leading Isgur-Wise functions in terms of $z$ around $z=0$, respectively.

\subsection{Bayesian model }

In EFTs, although in principle one 
can calculate relevant contributions to a particular observable up to any order, practical calculations must stay at a finite order due to the increasing number of free parameters at higher orders, which limited data cannot determine. The truncation errors induced by neglecting higher-order contributions dominate uncertainties in EFTs. 
To quantify the truncation errors of infinite EFT series and provide rigorous statistical errors for theoretical predictions, the Bayesian model~\cite{Furnstahl:2015rha,Melendez:2017phj,Melendez:2019izc} was proposed in recent years. This model has been successfully applied in EFTs studies,  such as relativistic chiral nuclear force~\cite{Lu:2021gsb}, meson-meson interaction~\cite{Zhai:2022ied} and antikaon-nucleon scattering~\cite{Lu:2022hwm}. In the HQET parameterization, the missing
higher-order contributions would influence the form factors and then shift the $|V_{cb}|$ value, potentially affecting the $V_{cb}$ puzzle. In this subsection, we explain how one can employ the Bayesian
model~\cite{Furnstahl:2015rha,Melendez:2017phj,Melendez:2019izc} to estimate the uncertainty of the extracted $|V_{cb}|$ value induced by neglecting higher-order `(beyond ${\cal O}(1/m_c^2)$)' contributions in HQET.

First, one should estimate the Bayesian uncertainties in the observables.
For any observable $X$, the EFT expansion is
\begin{equation}
X=X_{\mathrm{ref}} \sum_{n=0}^{\infty} c_n Q^n=X_{\mathrm{ref}} \left(\sum_{n=0}^{k} c_n Q^n +\Delta_k\right),\label{Eq:Byserror} 
\end{equation}
with 
\begin{equation}
\Delta_k=\sum_{n=k+1}^{\infty} c_n Q^n,
\end{equation}
where $\{c_n\}~(n=0,1,2,...)$ are dimensionless expansion coefficients. The product $X_{\mathrm{ref}}\Delta_k$ is the uncertainty term in the kth-order truncation of observable $X$, where $\Delta_k$ is a dimensionless parameter.
In the present work, the overall scale $X_{\rm ref}$ is defined as
\begin{equation}
X_{\mathrm{ref}}=
\operatorname{Max}\left\{\left|X^{\mathrm{LO}}\right|, \frac{\left|X^{\mathrm{LO}}-X^{\mathrm{NLO}}\right|}{Q}, \frac{\left|X^{\mathrm{NLO}}-X^{\mathrm{NNLO}}\right|}{Q^2}\right\},\label{Eq:Xref}
\end{equation}
with
\begin{equation}
Q=\operatorname{Max}\left\{\epsilon_a, \epsilon_b, \epsilon_c\right\},
\end{equation}
where $X_{\rm LO}$, $X_{\rm NLO}$, and $X_{\rm NNLO}$ are the prediction values for the observables in this study, calculated using the central values of three distinct parameter sets from Sect. \ref{secIIIB} obtained by fitting the available data up to LO, NLO, and NNLO respectively.

The dimensionless parameter $\Delta_k$  can be estimated by the known $\{c_n\}(n\leq k)$ and the integral of the posterior probability distribution function~(PDF) $\operatorname{pr}\left(\Delta \mid c_0, c_1, \ldots, c_k\right)$ given a certain degree of belief (DOB) $p\%$, which is 
\begin{equation}
p \%=\int_{-\Delta_k}^{\Delta_k} \operatorname{pr}\left(\Delta \mid c_0, c_1, \ldots, c_k\right) \mathrm{d} \Delta.
\label{Eq:pdf}
\end{equation}
The posterior PDF has the following form~\cite{Melendez:2017phj}, 
\begin{equation}
\operatorname{pr}\left(\Delta \mid c_0, c_1, \ldots, c_k\right)=\frac{\int_{1 / \bar{c}_{>}}^{1 / \bar{c}_{<}} d x x^{n_c} e^{-\left(\mathbf{c}_k^2+\Delta^2 / \bar{q}^2\right) x^2 / 2}}{\sqrt{2 \pi} \bar{q} \int_{1 / \bar{c}_{>}}^{1 / \bar{c}_{<}} d x x^{n_c-1} e^{-\mathbf{c}_k^2 x^2 / 2}}.\label{Eq:Pro}
\end{equation}
Equation~(\ref{Eq:Pro}) can be evaluated in terms of incomplete $\Gamma$ functions, which is
\begin{equation}
\operatorname{pr}\left(\Delta \mid c_0, c_1, \ldots, c_k\right)=\frac{1}{\sqrt{\pi \bar{q}^2 \bold{c}_k^2}}\left(\frac{\bold{c}_k^2}{\bold{c}_k^2+\Delta^2 / \bar{q}^2}\right)^{(n_c+1) / 2} \frac{\Gamma\left[\frac{n_c+1}{2}, \frac{1}{2 \bar{c}_{>}^2}\left(\bold{c}_k^2+\frac{\Delta^2}{\bar{q}^2}\right)\right]-\Gamma\left[\frac{n_c+1}{2}, \frac{1}{2 \bar{c}_{<}^2}\left(\bold{c}_k^2+\frac{\Delta^2}{\bar{q}^2}\right)\right]}{\Gamma\left[\frac{n_c}{2}, \frac{\bold{c}_k^2}{2 \bar{c}_{>}^2}\right]-\Gamma\left[\frac{n_c}{2}, \frac{\bold{c}_k^2}{2 \bar{c}_{<}^2}\right]},\label{Eq:ProIntegrated}
\end{equation}
where $\Gamma(s, x)=\int_x^{\infty} d t t^{s-1} e^{-t}$ and $\bold{c}_k^2 \equiv \sum_{i \in A} c_i^2, A \equiv\left\{i \in \mathbb{N}_0 \mid i \leq\right.$ $k \wedge i \neq m\}$ with $m$  corresponding to the term $c_m=1$. $n_c$ denotes the number of relevant known coefficients $\{c_n\}(n\leq k)$. However, the $c_m = 1$ term is excluded from this count, as it does not provide insight into the convergence pattern of the observables. The index $m$ in $c_m$ depends on the value of $X_{\rm ref}$ in Eq.~(\ref{Eq:Xref}). Although the Bayesian framework requires taking the asymptotic limit $h\to\infty$ for $\bar{q}^2 \equiv \sum_{i=k+1}^{k+h} Q^{2 i}$, we adopt a practical truncation at $h=10$ based on comprehensive numerical tests demonstrating complete convergence of the Bayesian uncertainties for observables at this cutoff value. As discussed in Ref.~\cite{Epelbaum:2019zqc}, Bayesian uncertainties are sensitive to the hyperparameter $\bar{c}_{<}$. In particular, taking the $\bar{c}_{<}\to0$ yields a $\delta$ function-like posterior $\operatorname{pr}\left(\Delta \mid c_0, c_1, \ldots, c_k\right)$, leading to an underestimation of Bayesian uncertainties. Nevertheless, we have checked that varying $\bar{c}_{<}$ within the range of $0.1–0.85$ does not affect the Bayesian error of $V_{cb}$ when the correction of the factor $\sqrt{\chi^2/\rm{d.o.f.}}$ is taken into account in our fits~(for detailed implementation procedures see section~\ref{secIIIA}). As a result, the robustness checks confirm that our choice for the $\bar{c}_{<}=0.5$ is numerically justified. For the other hyperparameter $\bar{c}_{>}\to\infty$, we have found that Bayesian uncertainties are insensitive to the parameter $\bar{c}_{>}$, justifying $\bar{c}_{>}=10$ used in our numerical analysis as an effective cutoff.

As stated above, for the truncation uncertainty or the Bayesian uncertainty $X_{\mathrm{ref}}\Delta_k$ of any observable $X$, $X_{\mathrm{ref}}$ can be estimated by Eq~.(\ref{Eq:Xref}) and $\Delta_k$ is determined a posteriori through Eq.~(\ref{Eq:pdf}) in which the DOB interval $p\%$ is given a prior. Statistically,  $p\%$  is the probability that the value of a certain observable $X$ falls within the $\left(X_{\rm c.v.}-X_{\mathrm{ref}}\Delta_k, X_{\rm c.v.}+X_{\mathrm{ref}}\Delta_k\right)$ interval. To ensure the reproducibility of our Bayesian results, we will provide a specific example of how to estimate the Bayesian error $X_{1\rm ref}\Delta_2$ of a given observable $\Delta\Gamma_1$ in the bin $\omega\in[1.00,1.05]$. Specifically, one uses the central values of parameters at LO, NLO, and NNLO in Sect. \ref{secIIIB} to obtain the prediction results of $\Delta\Gamma_{1\rm LO}$, $\Delta\Gamma_{1\rm NLO}$ and $\Delta\Gamma_{1\rm NNLO}$ respectively. Substituting these predictions and $Q={\rm Max}\{0.0716,0.0522,0.1807\}$ into Eq~.(\ref{Eq:Xref}), one gets $X_{1\rm ref}={\rm Max}\left\{0.0580, 0.0067, 0.0047\right\}$ . Then, the expansion coefficients  $c_0=\frac{\left|X^{\mathrm{LO}}\right|}{X_{1\rm ref}}=1$, $c_1=\frac{\left|X^{\mathrm{LO}}-X^{\mathrm{NLO}}\right|}{X_{1\rm ref}Q}=0.1160$ and $c_2=\frac{\left|X^{\mathrm{NLO}}-X^{\mathrm{NNLO}}\right|}{X_{1\rm ref}Q^2}=0.0804$ can be calculated by Eq.~(\ref{Eq:Byserror}). Using these known coefficients $c_1$ and $c_2$ as inputs and fixing $p\%=68\%$ in the equation $p \%=\int_{-\Delta_2}^{\Delta_2} \operatorname{pr}\left(\Delta \mid c_1, \ldots, c_2\right) \mathrm{d} \Delta$, one can determine the dimensionless parameter $\Delta_2=0.0048$. As a result, the Bayesian error of the observable $\Delta\Gamma_1$ is $X_{1\rm ref}\Delta_2=0.0003$. Note that the coefficient $c_0$ does not contribute to the PDF $\operatorname{pr}\left(\Delta \mid c_1, \ldots, c_2\right)$, i.e., $\bold{c}_k^2=c_1^2+c_2^2=0.0199$ in Eq.~(\ref{Eq:ProIntegrated}) because $c_0=1$ fails to provide meaningful information regarding the convergence behavior of the observables. The Bayesian uncertainties of other observables can be estimated following the same procedure.

Finally, at NNLO we re-fit the available data incorporating their Bayesian uncertainties. One expects that the central values of the unknown parameters remain nearly invariant while their uncertainties increase compared to the fit results without considering the Bayesian uncertainties. Therefore, the increased uncertainty for each parameter, such as $|V_{cb}|$, is the Bayesian uncertainty of that parameter.

\section{Results and discussions}

In this section, we first
perform the $|V_{cb}|$ fits in three different parameterizations for the form factors. Next, we study whether considering the truncation uncertainties induced by neglecting higher-order contributions in HQET can help solve the long-standing $V_{cb}$ puzzle. Finally, we predict some ratios for the $B\to D^*\ell\bar{\nu}_\ell$ decays, which can test the lepton-flavour universality in the SM.

\subsection{$|V_{cb}|$ determination in three different parameterizations}\label{secIIIA}

The previous studies~\cite{Grinstein:2017nlq,Bigi:2017njr,Bernlochner:2017jka,Jaiswal:2017rve,Bigi:2017jbd,Jung:2018lfu,Gambino:2019sif,Bordone:2019vic,Iguro:2020cpg,Bernlochner:2022ywh} have yielded two important findings: the $V_{cb}$ puzzle and the non-physical parameterization dependence for the $|V_{cb}|$ extraction. Recent Belle and Belle II  data~\cite{Belle:2023bwv,Belle-II:2023okj}, LQCD~\cite{FermilabLattice:2021cdg,Harrison:2023dzh,Aoki:2023qpa} and LCSR~\cite{Cui:2023jiw,Gubernari:2018wyi} studies provide an opportunity to revisit the above issues. To better understand the situation, i.e., whether the $V_{cb}$ puzzle is caused by different parameterizations, we study the $|V_{cb}|$ determinations in three different parameterizations: the CLN, BGL, and 
 HQET parameterizations. In the following numerical analysis, we take the PDG average~\cite{ParticleDataGroup:2024cfk} for the
meson masses. Other relevant inputs are collected in Appendix~\ref{App:Inputs}. Concerning the available experimental data, we ignore the lepton mass for the $|V_{cb}|$ extractions in the present work. As a result, the helicity amplitude $H_t$ will not contribute to the $B\to D^*(\to D\pi)\ell\bar{\nu}_\ell$ decays.  In other words, the form factors $R_0$, ${\cal F}_2$, and $A_0$ do not give any contribution.

In addition, it is well known that the experimental differential decay distributions, LQCD, and LCSR form factors all contribute to the determination of $|V_{cb}|$ from the four-body $B\to D^*(\to D\pi)\ell\bar{\nu}_{\ell}$ decays. To investigate the impact of different components on the $|V_{cb}|$ extraction, we investigate the following three cases:

\begin{itemize}

\item The Belle and Belle II data~\cite{Belle:2018ezy,Belle:2023bwv,Belle-II:2023okj}, the branching ratio of the $B\rightarrow D^*\ell\nu$~\cite{ParticleDataGroup:2024cfk} and the LQCD result $h_{A_1}(1)=0.906\pm0.013$~\cite{FermilabLattice:2014ysv} at $\omega=1$ are considered, denoted by the `${\rm Data}+h_{A_1}(1)+{\cal B}(\bar{B}\to D^{*+}\ell\bar{\nu}_\ell)$'. Thus, the total number of data fitted is 116. Note that the lattice input $h_{A_1}(1)$ at $\omega=1$ implies that our fitting can obtain a convergent $|V_{cb}|$ result.

\item Starting from case 1, we further consider the form factors at non-zero recoils~$(\omega>1)$ from different lattice collaborations~\cite{FermilabLattice:2021cdg,Harrison:2023dzh,Aoki:2023qpa}, which is referred to as `Data + LQCD.' More specifically, we included the form factors $h_V,h_{A_1},h_{A_2},h_{A_3}$ at $\omega=1.03, 1.10, 1.17$ and $q^2=0, q^2_{max}/4, 2q^2_{max}/4, 3q^2_{max}/4$ from the 
FNAL/MILC~\cite{FermilabLattice:2021cdg}, and the HPQCD ~\cite{Harrison:2023dzh} collabortions, and the form factors $g$, $f$, ${\cal F}_1$ at $\omega=1.025, 1.060, 1.100$ from the JLQCD Collaboration~\cite{Aoki:2023qpa}. The total number of data fitted
is 153. 

\item We additionally included the form factors $V$, $A_1$ at $\omega=1.50, 1.74, 1.98, 2.21$ and $V$, $A_1$, $A_2$ at $q^2=-3,-2,-1,0,1,2$ from the recent LCSR results~\cite{Gubernari:2018wyi,Cui:2023jiw}, compared to Case 2. This case is denoted as `Data + LQCD + LCSR'. A total of 179 data were fitted.
\end{itemize}

We emphasize that the HPQCD form factors~\cite{Harrison:2023dzh} at $\omega=1$ ($q^2=q^2_{max}$) are not used for two reasons. One is that the relation ${\cal F}_1(1)=m_B(1-r_{D^*})f(1)$ in the BGL parameterization can result in the interdependence and the singularity of the covariance matrix between the form factors as emphasized in \cite{FermilabLattice:2021cdg}. Another is that the other two parameterizations should adopt the same dataset for the purpose of comparison.  

\begin{table}[htb!]
\centering
\caption{Fit results using the CLN parameterization with experimental, LQCD, and LCSR data.\label{CLNFit}}
\begin{tabular}{ccccccccc}
\\ \hline \hline
~~~ CLN Fit~~~ & ~~~ Data +$h_{A_1}(0)=0.906$ + $\cal{B}$$(\bar{B}^0 \to D^{*+} \ell \bar{\nu}_{\ell})$~~~ &~~~ Data + LQCD~~~ &~~~ Data + LQCD + LCSR\\ \hline

$\chi^2/\rm{d.o.f.}$ & $137.44/112$ & $196.96/149$ & $219.17/175$  \\

$\rho_{D^*}^2$ & $1.194\pm0.031$ & $1.184\pm0.028$ & $1.147\pm0.026$  \\

$R_1(1)$ & $1.193\pm0.029$ & $1.225\pm0.024$ & $1.205\pm0.022$  \\

$R_2(1)$ & $0.858\pm0.020$ & $0.863\pm0.017$ & $0.884\pm0.016$  \\

$|V_{cb}|\times 10^{-3}$ & $39.99\pm0.79$ & $39.90\pm0.59$ & $39.69\pm0.56$  \\

 \hline \hline
\end{tabular}
\end{table}

\begin{table}[htb!]
\centering
\caption{Fit results using the BGL parameterization with experimental, LQCD, and LCSR data. Here, the ``Values I'' in Table~\ref{Inputs} is used for $\chi_{1^+}^T(0)$ and $\tilde{\chi}_{1^-}^T(0)$.}\label{BGLFit}
\begin{tabular}{ccccccccc}
\\ \hline \hline
~~~ BGL Fit~~~ & ~~~ Data +$h_{A_1}(0)=0.906$ + $\cal{B}$$(\bar{B}^0 \to D^{*+} \ell \bar{\nu}_{\ell})$ ~~~ &~~~ Data + LQCD~~~ &~~~ Data + LQCD + LCSR\\ \hline

$\chi^2/\rm{d.o.f.}$ & $135.03/107$ & $164.70/144$ & $188.69/170$  \\

$a_0^g$ & $0.0240\pm0.0101$ & $0.0270\pm0.0004$ & $0.0265\pm0.0005$  \\

$a_1^g$ & $0.044\pm0.365$ & $-0.068\pm0.0021$ & $-0.106\pm0.022$  \\

$a_2^g$ & {$-0.999^{+1.999}_{-0.001}$} & $-0.9973\pm0.0002$ & $0.424\pm0.324$  \\

$a_0^f$ & $0.0122\pm0.0002$ & $0.0122\pm0.0001$ & $0.0122\pm0.0001$ \\

$a_1^f$ & $0.033\pm0.040$ & $0.017\pm0.006$ & $0.008\pm0.005$  \\

$a_2^f$ & {$-0.904^{+1.904}_{-0.096}$} & $-0.436\pm0.155$ & $-0.121\pm0.109$  \\

$a_1^{{\cal{F}}_1}$ & $0.002\pm0.002$ & $0.0007\pm0.0011$ & $0.001\pm0.001$  \\

$a_2^{{\cal{F}}_1}$ & $-0.016\pm0.031$ & $0.008\pm0.021$ & $-0.007\pm0.021$  \\

$|V_{cb}|\times 10^{-3}$ & $39.48\pm0.94$ & $39.91\pm0.56$ & $39.90\pm0.55$  \\

 \hline \hline
\end{tabular}
\end{table}

\begin{table}[h!]
\centering
\caption{Fit results using the HQET~(2/1/0) parameterization with experimental, LQCD, and LCSR data.}\label{HQETFit}
\begin{tabular}{ccccccccc}
\\ \hline \hline
~~~ HQET~(2/1/0) Fit~~~ & ~~~ Data +$h_{A_1}(0)=0.906$ + $\cal{B}$$(\bar{B}^0 \to D^{*+} \ell \bar{\nu}_{\ell})$~~~ &~~~ Data + LQCD~~~ &~~~ Data + LQCD + LCSR\\ \hline
$\chi^2/\rm{d.o.f.}$ & $136.78/104$ & $208.64/141$ & $222.80/167$  \\

$|V_{cb}|\times10^3$ & $39.89\pm0.86$ & $39.42\pm0.63$ & $39.48\pm0.59$  \\

$\xi^{(1)}$ & $-1.294\pm0.069$ & $-1.233\pm0.043$ & $-1.234\pm0.038$  \\

$\xi^{(2)}$ & $2.052\pm0.314$ & $1.649\pm0.184$ & $1.736\pm0.137$  \\

$\hat{\chi}_2^{(0)}$ & $-0.060\pm0.023$ & $-0.059\pm0.024$ & $-0.060\pm0.023$  \\

$\hat{\chi}_2^{(1)}$ & $0.0004\pm0.0229$ & $0.008\pm0.024$ & $0.005\pm0.023$  \\

$\hat{\chi}_3^{(0)}(\rm{fixed})$ & $0$ & $0$ & $0$  \\

$\hat{\chi}_3^{(1)}$ & $0.035\pm0.029$ & $0.036\pm0.030$ & $0.036\pm0.029$  \\

$\eta^{(0)}$ & $0.612\pm0.132$ & $0.553\pm0.139$ & $0.571\pm0.132$  \\

$\eta^{(1)}$ & $0.041\pm0.034$ & $0.054\pm0.036$ & $0.052\pm0.035$  \\

$\hat{\ell}_2^{(0)}$ & $-1.964\pm0.456$ & $-2.001\pm0.259$ & $-1.985\pm0.245$  \\

$\hat{\ell}_3^{(0)}$ & $-3.854\pm97.532$ & $1.113\pm1.742$ & $0.770\pm1.638$  \\

$\hat{\ell}_5^{(0)}$ & $2.810\pm1.040$ & $1.416\pm0.897$ & $1.895\pm0.828$  \\

$\hat{\ell}_6^{(0)}$ & $1.732\pm43.766$ & $2.571\pm1.147$ & $3.109\pm1.061$  \\

 \hline \hline
\end{tabular}
\end{table}

To ensure that the  parameter errors obtained in our fits, especially for $V_{cb}$, are not underestimated, we apply a standard scaling correction to all fitting results on the $|V_{cb}|$ determination following the PDG prescription~\cite{ParticleDataGroup:2024cfk}. This procedure involves multiplying the uncertainty of fitted parameter (such as $V_{cb}$) by the scale factor $\sqrt{\chi^2/\rm{d.o.f.}}$, with the $\rm{d.o.f.}$ representing the degree of freedom, while leaving the central value of that parameter unaffected. In this way, we present all $V_{cb}$-related fitting results in Tables~\ref{CLNFit}, \ref{BGLFit}, \ref{HQETFit}, \ref{BayesFit}, \ref{BGLFitLQCD1}, and \ref{BGLFitLQCD2}. These results show that the $|V_{cb}|$ values extracted with the three different parameterizations are consistent with each other: 
\begin{eqnarray}
&&|V_{cb}|_{\rm CLN}=(39.69\pm0.56)\times10^{-3},\nonumber\\
&&|V_{cb}|_{\rm BGL}=(39.90\pm0.55)\times10^{-3},\nonumber\\
&&|V_{cb}|_{\rm HQET~(2/1/0)}=(39.48\pm0.59)\times10^{-3}.
\end{eqnarray}
and there is no parameterization dependence. This is because the Belle 2023 data~\cite{Belle:2023bwv} are incompatible with the Belle 2017 data~\cite{Belle:2017rcc}. In the present analysis, we use the Belle 2023 data of higher statistics, which supersede those of Belle 2017. It is shown that our exclusive determinations for the $|V_{cb}|$ deviate from the inclusive results~\cite{Bernlochner:2022ucr,Finauri:2023kte} with the significance of $2\sigma\sim3\sigma$. As a result, the $V_{cb}$ puzzle has not been resolved.

The extraction of the $|V_{cb}|$ value in each parameterization is carried out in the three different cases to investigate the impact of the experimental data, the LQCD and LCSR form factors on the $V_{cb}$ issue. In the BGL parameterization, comparing the three scenarios, we found the LQCD data had a greater influence on the central values of the $a_1^g,a_1^f,a_2^f,a_2^{{\cal F}_1}$ and the LCSR results constrained $a_1^g,a_2^g,a_1^f,a_2^f$ . The LQCD data led to smaller uncertainties for the fitting parameters, and the LCSR inputs did not affect them. Our fitting procedure has explicitly incorporated unitarity constraints. We note that the parameters $a_2^g$ and $a_2^f$ cannot be effectively constrained by the experimental data, LQCD, or LCSR individually. The same phenomenon is reported in Refs.~\cite{Bigi:2017njr,Bigi:2017jbd,Gambino:2019sif,Kapoor:2024ufg}. To rigorously satisfy unitarity conditions in our fitting, we have implemented carefully designed non-Gaussian probability distributions for these two parameters~(e.g., Table~\ref{BGLFit}). Compared to the latest $|V_{cb}|$ determination~\cite{Bordone:2024weh,Kapoor:2024ufg} in the BGL parameterization, the center value and uncertainty of our results are relatively lower. The reason is that in Ref.~\cite{Bordone:2024weh,Kapoor:2024ufg} the data from LCSR and Belle 2018 are not considered. In addition, we also perform a systematic analysis of how the theoretical inputs for $\chi_{1^+}^T(0)$ and $\tilde{\chi}_{1^-}^T(0)$- computed using perturbative QCD~\cite{Bigi:2017njr} and LQCD~\cite{Harrison:2024iad,Martinelli:2021frl} approaches-influence the BGL fit results. The relevant results are presented in Tables~\ref{BGLFit}, \ref{BGLFitLQCD1}, and \ref{BGLFitLQCD2}. One finds that the $|V_{cb}|$ values obtained through the BGL parameterization using three distinct theoretical inputs in Table~\ref{Inputs} are consistent with each other at $1\sigma$ confidence level once constraints from the LQCD data are taken into account. Comparative analysis shows that the correlations differ somewhat between Table~\ref{BGLCorr} and Tables~\ref{BGLCorrII} and \ref{BGLCorrIII}. We also note that the accuracy of our $|V_{cb}|$ value in the CLN parameterization is twice that of the previous work~\cite{Bigi:2017njr}. Under the HQET parameterization, the $|V_{cb}|$ value we obtained is at the same level of accuracy as that of a recent work~\cite{Bernlochner:2022ywh} in considering the combined constraint of the $B\to D^*$ and $B\to D$ processes. The NNLO parameters in the HQET, i.e., $\hat{\ell}_3^{(0)}$, $\hat{\ell}_5^{(0)}$, and $\hat{\ell}_6^{(0)}$, are sensitive to the constraints from LQCD and LCSR. All of these indicate the importance of the new experimental data and LQCD and LCSR results on the $|V_{cb}|$ extraction. Therefore, more precise experimental data, LQCD, and LCSR calculations are needed, which can help better understand the nature of the $V_{cb}$ puzzle.

\subsection{Estimation of theoretical uncertainties beyond the NNLO contribution in the HQET}\label{secIIIB}

We have performed an analysis of the $B\to D^*$ form factors within the HQET up to ${\cal O}(1/m_c^2)$ order for the $V_{cb}$ study. The $V_{cb}$ puzzle persists. At NNLO in the HQET, we only consider the contribution from the $1/m_c^2$ expansion term because other ${\cal{O}}(1/m_bm_c,1/m_b^2,\alpha_s/m_{b,c})$ corrections are not sensitive to the available data and can be safely neglected~\cite{Bernlochner:2022ywh}.  Because more parameters are involved at higher orders and current data are limited, we have neglected higher-order terms beyond the ${\cal O}(1/m_c^2)$ order that they can change the shape of the form factors, leading to a shift of the $|V_{cb}|$ value. In the following, we apply the state-of-the-art Bayesian approach proposed in Refs.~\cite{Furnstahl:2015rha,Melendez:2017phj,Melendez:2019izc} to estimate truncation uncertainties originating from the missing
higher-order contributions to the $|V_{cb}|$ determination.

\begin{table}[htb!]
\centering
\caption{Fit results using the HQET~(2/1/0) parameterizations at LO, NLO, and NNLO, respectively, in Case 3. The last column shows the Bayesian fit results up to NNLO.}\label{BayesFit}
\begin{tabular}{ccccccccc}
\\ \hline \hline
~~~ HQET~(2/1/0) Fit~~~ & ~~~ LO~~~ &~~~ NLO~~~&~~~ NNLO~~~&~~~ Bayesian results\\ \hline

$\chi^2/\rm{d.o.f.}$ & $586.29/176$ & $324.68/171$ &  $222.80/167$& $187.17/167$  \\

$|V_{cb}|\times10^3$ & $36.57\pm0.69$ & $36.69\pm0.52$ & $39.48\pm0.59$& $39.63 \pm 0.59$ \\

$\xi^{(1)}$ & $-1.031\pm0.055$ & $-1.271\pm0.041$ & $-1.234\pm0.038$& $-1.264 \pm 0.044$  \\

$\xi^{(2)}$ & $1.353\pm0.219$ & $1.742\pm0.152$ & $1.736\pm0.137$& $1.904\pm 0.161$  \\

$\hat{\chi}_2^{(0)}$ & $0$ & $-0.060\pm0.028$ &$-0.060\pm0.023$& $-0.060\pm0.021$  \\

$\hat{\chi}_2^{(1)}$ & $0$ & $0.002\pm0.028$ &$0.005\pm0.023$&  $0.005 \pm 0.021$  \\

$\hat{\chi}_3^{(0)}(\rm{fixed})$ & $0$ & $0$ & $0$ & $0$  \\

$\hat{\chi}_3^{(1)}$ & $0$ & $0.035\pm0.033$ & $0.036\pm0.029$ &  $0.035\pm0.026$  \\

$\eta^{(0)}$ & $0$ & $0.471\pm0.055$ &  $0.571\pm0.132$ &  $0.580\pm0.121$ \\

$\eta^{(1)}$ & $0$ & $0.050\pm0.041$ &  $0.052\pm0.035$ & $0.049\pm0.032$  \\

$\hat{\ell}_2^{(0)}$ & $0$ & $0$ &  $-1.985\pm0.245$ &  $-1.934\pm0.278$\\

$\hat{\ell}_3^{(0)}$ & $0$ & $0$ & $0.770\pm1.638$ &  $-0.025\pm1.586$ \\

$\hat{\ell}_5^{(0)}$ & $0$ & $0$ & $1.895\pm0.828$ &  $2.137\pm0.820$  \\

$\hat{\ell}_6^{(0)}$ & $0$ & $0$ & $3.109\pm1.061$ &  $3.066\pm1.021$  \\

 \hline \hline
\end{tabular}
\end{table}

To investigate the influence of the PDG scaling procedure on the Bayesian uncertainty for $|V_{cb}|$, we apply the scaling factor $\sqrt{\chi^2/\rm{d.o.f.}}$ to the uncertainties of $|V_{cb}|$ obtained from the LO, NLO, NNLO, and Bayesian fit in the HQET~(2/1/0) parameterization. The scaled results are presented in Table~\ref{BayesFit}. One can see that the Bayesian uncertainty of $|V_{cb}|$ is close to 0. As a result, we conclude that higher-order effects in the HQET are completely negligible. Following the procedures of the Bayesian framework, our analysis obtained a bad fit result, i.e., $|V_{cb}|=(36.69\pm0.52)\times10^{-3}$,
at NLO in the HQET, which is very different from the $|V_{cb}|=(39.3\pm1.0)\times10^{-3}$ in Ref.~\cite{Bernlochner:2017jka}. Apart from the different datasets used, the main reason for the significant difference is that Ref.~\cite{Bernlochner:2017jka} re-scaled the $B\rightarrow D$ and $B\rightarrow D^*$ form factors in the fit by ${\cal G}(1)_{\rm {LQCD}}/{\cal G}(1)$ and ${\cal F}(1)_{\rm{LQCD}}/{\cal F}(1)$, such that the rates at $\omega=1$ agree with the LQCD predictions. In contrast, we did not re-scale the form factors. More specifically, the value ${\cal F}(1)=h_{A_1}(1)=1+\alpha_sC_{A_1}(1)=0.966$ we obtained at NLO is about a $5\%$ upward shift, compared with that from the LQCD result~\cite{Bernlochner:2017jka}. This results in a $5\%$ downward shift for the $V_{cb}$ value without rescaling the form factors. Nevertheless, Table~\ref{BayesFit} shows that considering the NNLO corrections can significantly improve the $|V_{cb}|$ result at NLO. This indicates that it is essential to consider the NNLO contributions of the form factors in the HQET. 

Notably, a prior study~\cite{Crivellin:2014zpa} based on total rates alone has shown that the effects of new physics cannot explain the $V_{cb}$ puzzle. However, the inclusion of angular distributions of $B\to D^*\ell\bar{\nu}_\ell$  may change this picture as the shape of these observables is sensitive to new physics~\cite{Jung:2018lfu}. As a result, the possibility of new physics accounting for the $|V_{cb}|$ deviation still cannot be ruled out and the $V_{cb}$ analyses considering new physics contributions have been performed in some works~\cite{Jung:2018lfu,Iguro:2020cpg,Kapoor:2024ufg,Colangelo:2016ymy,Colangelo:2018cnj,Huang:2021fuc}.

\subsection{SM predictions of observables for the $B\to D^*\ell\bar{\nu}_\ell$ decay}

\begin{figure}[htb!]
\begin{tabular}{cc}
  \includegraphics[width=7cm,height=4.5cm]{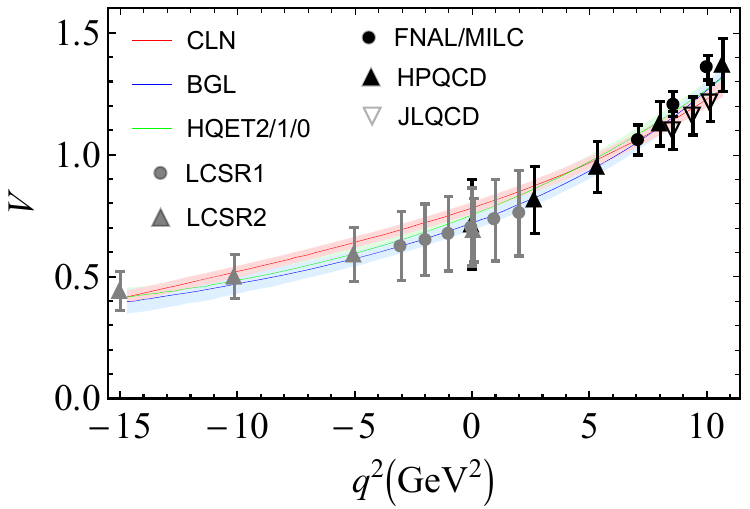}~~~\includegraphics[width=7cm,height=4.5cm]{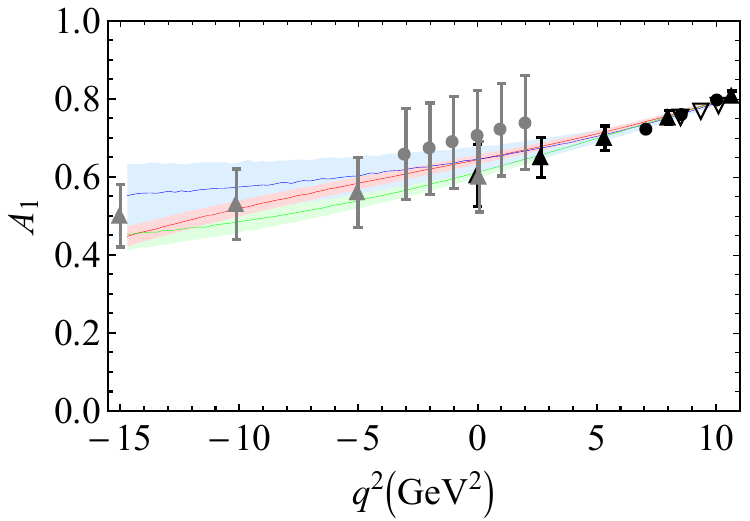}\\
  \includegraphics[width=7cm,height=4.5cm]{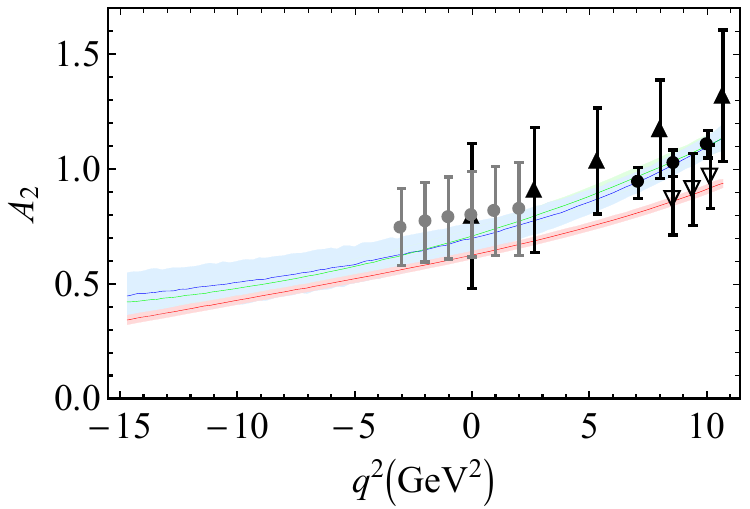}~~~\includegraphics[width=7cm,height=4.5cm]{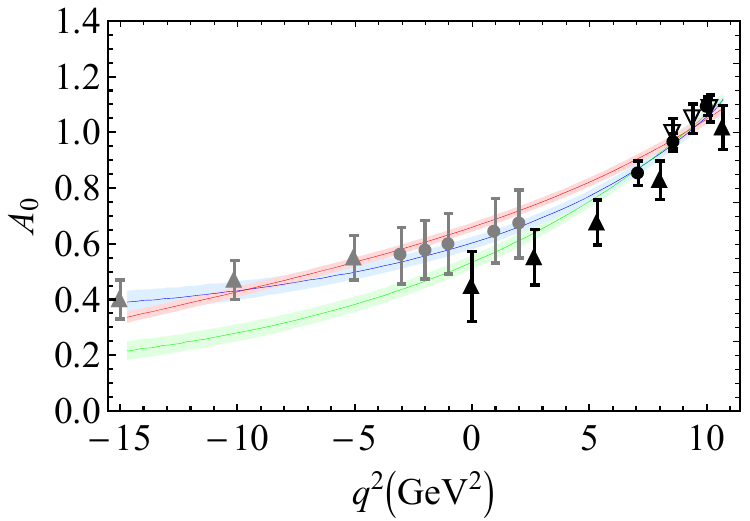}\\

\end{tabular}
\caption{Form factors $A_0(q^2)$, $A_1(q^2)$, $A_2(q^2)$, and $V(q^2)$ dependence on $q^2$ for the three different form-factor parametrizations: CLN (red bands), BGL (blue bands), and HQET~(2/1/0) (green bands) using the
best-fit results from the combined LQCD+LCSR analysis as inputs. The shaded bands show the regions with the $1\sigma$ upper and lower limits of the form factors. The black and gray points with error bars are the LQCD and LCSR data used in our fits, respectively. In the figure, the error bars of $A_1(q^2)$  from FNAL/MILC and JLQCD are too small to be seen.}\label{Fig:Obsq2}
\end{figure}

In this subsection, we focus on three quantities: $R_{D^*}$, $P_\tau^{D^*}$, and $F_L^{D^*}$, whose definitions can be found in Ref.~\cite{Shi:2019gxi}. It should be noted that the three observables are insensitive to hadronic uncertainties and do not depend on $V_{cb}$. As the experimental data may include the NP effects, one cannot use the parameters in Tables~\ref{CLNFit} to \ref{HQETFit} as inputs to predict the $R_{D^*}$, $P_\tau^{D^*}$, and $F_L^{D^*}$ in the SM. Nevertheless, the inputs related to the SM predictions can be determined via fitting the available LQCD and LCSR data. Here, we have additionally incorporated the following LQCD and LCSR data constraining the CLN coefficient $R_0(1)$, as well as the BGL coefficients $a_0^{{\cal{F}}_2}$ and $a_1^{{\cal{F}}_2}$: ${\cal F}_2$ from the JLQCD Collaboration at $\omega=1.025, 1.060, 1.100$ and $A_0$ at $q^2=1.74, 1.98, 2.21$  and $q^2=-3, -2, -1, 1, 2$ from the LCSR results~\cite{Gubernari:2018wyi,Cui:2023jiw}. In Tables~\ref{CLNFitLQCDLCSR} to \ref{HQETFitLQCDLCSR}, we present the fitting results for the CLN, BGL, and HQET~(2/1/0) parameterizations, respectively. One can see that the parameters extracted solely from the LCSR data exhibit larger uncertainties than those obtained from the LQCD or combined LQCD+LCSR fits, due to the larger theoretical errors inherent in the LCSR calculations. From the combined LQCD+LCSR analysis, the smaller $\chi^2/{\rm d.o.f.}$ indicates that the BGL and HQET parameterization fits agree more closely with the LQCD and LCSR data than the CLN fit. These best-fit results in Tables~\ref{CLNFitLQCDLCSR} to \ref{HQETFitLQCDLCSR} allow us to predict the observables $R_{D^*}$, $P_\tau^{D^*}$, and $F_L^{D^*}$ in the SM. On the other hand, in Fig.~\ref{Fig:Obsq2}, we plot the relevant form factors as functions of $q^2$ for the CLN, BGL, and HQET parameterizations. It is demonstrated that the form factors determining the three observables at the $q^2>0$ region are consistent with each other with the the preferred BGL and HQET parameterizations when utilizing the best-fit results from the combined LQCD+LCSR analysis as inputs. Our predictions and those in a recent study~\cite{Martinelli:2024bov} for form factors are compatible within errors. As a result, we will choose the fit parameters of the HQET~(2/1/0) to predict the $R_{D^*}$, $P_\tau^{D^*}$, and $F_L^{D^*}$.
\begin{table}[htb!]
\centering
\caption{Fit results using the CLN parameterization with the LQCD and LCSR data only.}\label{CLNFitLQCDLCSR}
\begin{tabular}{ccccccccc}
\\ \hline \hline
~~~ CLN Fit~~~ & ~~~ LQCD~~~ &~~~ LCSR~~~ &~~~LQCD + LCSR\\ \hline

$\chi^2/\rm{d.o.f.}$ & $51.17/37$ & $5.97/30$ & $94.42/71$  \\

$\rho_{D^*}^2$ & $1.16\pm0.06$ & $0.91\pm0.10$ & $1.02\pm0.03$  \\

$R_1(1)$ & $1.33\pm0.04$ & $1.06\pm0.06$ & $1.26\pm0.03$  \\

$R_2(1)$ & $0.89\pm0.02$ & $1.10\pm0.06$ & $0.93\pm0.02$  \\

$R_0(1)$ & $1.11\pm0.02$ & $0.95\pm0.05$ & $1.08\pm0.02$  \\

 \hline \hline
\end{tabular}
\end{table}
\begin{table}[htb!]
\centering
\caption{Fit results using the BGL parameterization with the LQCD and LCSR data only. Here, the ``Values I'' in Table~\ref{Inputs} is used for $\chi_{1^+}^T(0)$, $\tilde{\chi}_{1^-}^T(0)$, and $\tilde{\chi}_{1+}^L(0)$.}\label{BGLFitLQCDLCSR}
\begin{tabular}{ccccccccc}
\\ \hline \hline
~~~ BGL Fit~~~ & ~~~ LQCD~~~ &~~~ LCSR~~~ &~~~LQCD + LCSR\\ \hline

$\chi^2/\rm{d.o.f.}$ & $17.70/31$ & $1.36/24$ & $29.15/65$  \\

$a_0^g$ & $0.028\pm0.001$ & $0.024\pm0.008$ & $0.027\pm0.001$  \\

$a_1^g$ & $-0.04\pm0.03$ & $-0.06\pm0.24$ & $-0.06\pm0.02$  \\

$a_2^g$ & $-0.99^{+1.96}_{-0.01}$ & $0.52^{+0.48}_{-1.52}$ & $0.07\pm0.34$  \\

$a_0^f$ & $0.0121\pm0.0001$ & $0.0139\pm0.0039$ & $0.0122\pm0.0001$ \\

$a_1^f$ & $0.013\pm0.007$ & $-0.01\pm0.11$ & $0.012\pm0.006$  \\

$a_2^f$ & $-0.14\pm0.29$ & $0.02\pm0.73$ & $0.07\pm0.11$  \\

$a_1^{{\cal{F}}_1}$ & $-0.003\pm0.002$ & $0.005\pm0.019$ & $-0.003\pm0.002$  \\

$a_2^{{\cal{F}}_1}$ & $-0.03\pm0.062$ & $-0.19\pm0.16$ & $0.03\pm0.02$  \\

$a_0^{{\cal{F}}_2}$ & $0.049\pm0.001$ & $0.046\pm0.014$ & $0.049\pm0.001$  \\

$a_1^{{\cal{F}}_2}$ & $-0.19\pm0.04$ & $-0.10\pm0.39$ & $-0.19\pm0.04$  \\

 \hline \hline
\end{tabular}
\end{table}

\begin{table}[htb!]
\centering
\caption{Fit results using the HQET (2/1/0) parameterization with the LQCD and LCSR data only.}\label{HQETFitLQCDLCSR}
\begin{tabular}{ccccccccc}
\\ \hline \hline
~~~ HQET (2/1/0) Fit~~~ & ~~~ LQCD~~~ &~~~ LCSR~~~ &~~~LQCD + LCSR\\ \hline

$\chi^2/\rm{d.o.f.}$ & $20.20/26$ & $2.36/15$ & $41.76/52$  \\

$\xi^{(1)}$ & $-1.26\pm0.05$ & $-0.99\pm0.31$ & $-1.19\pm0.05$  \\

$\xi^{(2)}$ & $1.39\pm0.15$ & $1.21\pm0.80$ & $1.56\pm0.14$  \\

$\hat{\chi}_2^{(0)}$ & $-0.06\pm0.02$ & $-0.06\pm0.02$ & $-0.06\pm0.02$  \\

$\hat{\chi}_2^{(1)}$ & $0.002\pm0.020$ & $0.00\pm0.02$ & $-0.002\pm0.020$  \\

$\hat{\chi}_3^{(0)}(\rm{fixed})$ & $0$ & $0$ & $0$  \\

$\hat{\chi}_3^{(1)}$ & $0.035\pm0.025$ & $0.035\pm0.025$ & $0.036\pm0.025$  \\

$\eta^{(0)}$ & $0.61\pm0.12$ & $0.62\pm0.11$ & $0.62\pm0.11$  \\

$\eta^{(1)}$ & $0.04\pm0.03$ & $0.04\pm0.03$ & $0.05\pm0.03$  \\

$\hat{\ell}_2^{(0)}$ & $-2.17\pm0.22$ & $-1.69\pm3.88$ & $-2.13\pm0.21$  \\

$\hat{\ell}_3^{(0)}$ & $-6.56\pm2.04$ & $-5.15\pm12.77$ & $-7.13\pm2.01$  \\

$\hat{\ell}_5^{(0)}$ & $-1.55\pm1.04$ & $8.12\pm2.41$ & $-0.35\pm1.00$  \\

$\hat{\ell}_6^{(0)}$ & $2.81\pm1.10$ & $9.63\pm5.11$ & $3.75\pm1.07$  \\

 \hline \hline
\end{tabular}
\end{table}
\begin{table}[htb!]
\centering
\caption{SM predictions for binned observables integrated over
the whole kinematic region in the LQCD, LSCR, and LQCD+LCSR scenarios.}\label{SMPre}
\begin{tabular}{ccccc}
\hline \hline
~~~Observables~~~ & ~~~LQCD~~~ & ~~~LCSR~~~ & ~~~LQCD + LCSR ~~~&~~~ Exp. \\ \hline
$R_{D^*}$ & $0.270\pm0.005$    &$0.264^{+0.031}_{-0.034}$ & $0.262\pm0.003$ &$0.287\pm0.012$~\cite{hflav2024}\\

$P_{\tau}^{D^*}$ & $-0.523\pm0.005$  & $-0.503^{+0.191}_{-0.130}$ &  $-0.522\pm0.005$  & $-0.38\pm0.51_{-0.16}^{+0.21}$~\cite{Belle:2016dyj} \\

$F_L^{D^*}$ & $0.425\pm0.007$  &$0.440^{+0.150}_{-0.153}$ &  $0.425\pm0.007$&
\begin{tabular}{@{}c@{}} 
$0.43\pm0.06\pm0.03$~\cite{LHCb:2023ssl} \\
$0.60\pm0.08\pm0.04$~\cite{Belle:2019ewo}
\end{tabular} \\

\hline \hline
\end{tabular}
\end{table}
Table~\ref{SMPre} summarizes the SM predictions for the three binned observables integrated over the whole kinematic region in the LQCD, LSCR, and LQCD+LCSR scenarios. One finds that $P_\tau^{D^*}$ and $F_L^{D^*}$ are insensitive to the LSCR data because of their large theoretical uncertainties. In the following, we employ the results in the LQCD+LCSR scenario as our theoretical baseline. Our predictions for the $R_{D^*}$, $P_\tau^{D^*}$, and $F_L^{D^*}$ differ by $1.5\sigma\sim3\sigma$ from the previous HQET-based calculations~\cite{Bordone:2019vic,Iguro:2020cpg,Bernlochner:2022ywh,Huang:2018nnq}. This discrepancy stems from differences in the HQET parameter determination, as Refs.~\cite{Bordone:2019vic,Iguro:2020cpg,Bernlochner:2022ywh,Huang:2018nnq} incorporate experimental data in their analyses. On the other hand, one can see that the predicted $R_{D^*}$ is in tension with the current HFLAV result~\cite{hflav2024} by approximately $1.8\sigma$. Besides, our prediction for $F_L^{D^*}$ is in good agreement with the recent LHCb measurement~\cite{LHCb:2023ssl} but deviates from the previous Belle data~\cite{Belle:2019ewo} with a significance of $\sim1.9\sigma$. These results indicate lepton-flavor universality violation in the $b\to c\tau\nu$ transitions, which future experiments should verify. Future high-precision experiments can also test our prediction for the $P_\tau^{D^*}$.
\section{summary}

In this work, we investigated the $V_{cb}$ puzzle with the CLN, BGL, and HQET parameterizations by using the latest Belle and Belle II results on the decay distributions in $B\rightarrow D^* (\rightarrow D\pi)\ell\bar{\nu}_\ell$ and considering constraints from recent LQCD simulations and light-cone sum rule results. We obtained $|V_{cb}|_{\rm CLN} = ( 39.69\pm0.56)\times10^{-3}, |V_{cb}|_{\rm BGL} = (39.90\pm0.55)\times10^{-3} $ and $ |V_{cb}|_{\rm HQET} = (39.48\pm0.59)\times10^{-3}$. These three parameterizations lead to consistent results for $|V_{cb}|$ but still show $\sim2$ and $\sim3$ standard deviations compared to the latest inclusive determinations $|V_{cb}|_{\rm incl} = (41.69\pm0.63)\times10^{-3}$~\cite{Bernlochner:2022ucr} and $|V_{cb}|_{\rm {incl}}= (41.97\pm0.48)\times 10^{-3}$~\cite{Finauri:2023kte}. Next, using the Bayesian method, we quantitatively analyzed the impact of higher-order terms in the HQET on the uncertainty of $V_{cb}$. We obtained a Bayesian uncertainty for $|V_{cb}|$ $\sim0$, which is not enough to fill the $|V_{cb}|$ gap between the exclusive and inclusive modes. Therefore, we conclude that more experimental data or new physics effects are needed to understand the $V_{cb}$ puzzle. 

In addition, we provided SM predictions for $B\to D^*\ell\bar{\nu}_\ell$ observables using the best-fit results from the combined LQCD+LCSR analysis as inputs. Our result for $R_{D^*}$ deviates from that in the current HFLAV report~\cite{hflav2024} with a significance of $\sim1.8\sigma$. Furthermore, our prediction for $F_L^{D^*}$ is consistent with the recent LHCb measurement~\cite{LHCb:2023ssl} but in tension with the previous Belle data~\cite{Belle:2019ewo} with a significance of $\sim1.9\sigma$. These results suggest that the lepton-flavour universality in the $b\to c\tau\nu$ transitions may be violated, which should be verified by future experiments.

As well known, the $|V_{cb}|$ value is sensitive to the slope of the form factors. In the present work, we do not consider the constraint of the $B\to D\ell\bar{\nu}_\ell$ processes on the form factors in the HQET parameterization.  Further improvements are necessary. The theoretical uncertainties and new physics contributions should all be considered to solve the $V_{cb}$ puzzle. We leave these for future work.

\section{Acknowledgments}
This work is partly supported by the National Natural Science Foundation of China under Grants No. 12405091 and the Natural Science Foundation of Guangxi province under Grant No. 2025GXNSFBA069314. Li-Sheng Geng acknowledges support from the National Key R\&D Program of China under Grant No. 2023YFA1606703 and the National Natural Science Foundation of China under Grant No. 12435007. Yi Zhang acknowledges support from the National Natural Science Foundation of China under Grants No. 12347182 and the Project funded by China Postdoctoral Science Foundation No. 2023M740190.

\newpage
\appendix
\section{Relevant theoretical inputs for the $|V_{cb}|$ extraction}\label{App:Inputs}
Relevant theoretical inputs for the $|V_{cb}|$ extraction are collected in Table~\ref{Inputs}.
\begin{table}[htb!]
\centering
\caption{Relevant theoretical inputs~\cite{Bigi:2017njr,ParticleDataGroup:2024cfk,Harrison:2024iad,Martinelli:2021frl} for the $|V_{cb}|$ extraction.\label{Inputs}}
\begin{tabular}{ccccccccc}
\\ \hline \hline
Common inputs & Values\\ \hline
$\tau_{B^0}$ & $1.519\times 10^{-12}$s  \\

$\cal{B}$$(D^{*+} \to D^0 \pi^+)$ & 0.677  \\

$\cal{B}$$(D^0 \to K^- \pi^+)$ &  0.03947  \\

$\cal{B}$$(\bar{B}^0 \to D^{*+} \ell \bar{\nu}_{\ell})$ &  ($4.97\pm 0.12$)\%  
\\
$G_F$ & $1.1663787\times 10^{-5}$($\rm GeV^{-2}$)  \\
 \hline
 BGL inputs\\
 \hline
Vector $B_c^*$ mass $1^-$ for $g$  & $6.329$(GeV)  \\
 & $6.920$(GeV)  \\
 & $7.020$(GeV)  \\
 & $7.280$(GeV)  \\

Axial vector $B_c^*$ mass $1^+$ for 
 $f$,${\cal F}_1$ & $6.739$(GeV)  \\
 & $6.750$(GeV)  \\
 & $7.145$(GeV)  \\
 & $7.150$(GeV)  \\
$n_I$ & $2.6$  \\

Values I & \begin{tabular}{@{}c@{}} 
$\tilde{\chi}_{1^-}^T(0)=5.131 \times 10^{-4}$($\rm GeV^{-2}$)~\cite{Bigi:2017njr,Bigi:2017jbd}  \\

$\chi_{1^+}^T(0)=3.894 \times 10^{-4}$($\rm GeV^{-2}$)~\cite{Bigi:2017njr,Bigi:2017jbd}  \\

$\tilde{\chi}_{1^+}^L(0)=1.9421 \times 10^{-2}$~\cite{Bigi:2017njr,Bigi:2017jbd}  \\
\end{tabular} \\ 

Values II & \begin{tabular}{@{}c@{}} 
$\tilde{\chi}_{1^-}^T(0)=6.55 \times 10^{-4}$($\rm GeV^{-2}$)~\cite{Harrison:2024iad}  \\

$\chi_{1^+}^T(0)=4.06 \times 10^{-4}$($\rm GeV^{-2}$)~\cite{Harrison:2024iad} \\
\end{tabular} \\ 

Values III & \begin{tabular}{@{}c@{}} 
$\tilde{\chi}_{1^-}^T(0)=5.84 \times 10^{-4}$($\rm GeV^{-2}$)~\cite{Martinelli:2021frl}   \\

$\chi_{1^+}^T(0)=4.69 \times 10^{-4}$($\rm GeV^{-2}$)~\cite{Martinelli:2021frl} \\ 
\end{tabular} \\ 

 \hline \hline
\end{tabular}
\end{table}

\section{Corrections of leading Isgur-Wise function in the HQETparametrization}\label{App:HQETcorr}

The $\alpha_s$ corrections are given as
\begin{eqnarray}
&&\delta \hat{h}_{V, \alpha_s}= \frac{1}{6 z_{c b}\left(\omega-\omega_{c b}\right)}\left[4 z_{c b}\left(\omega-\omega_{c b}\right) \Omega_\omega(\omega)+2(\omega+1)\left((3 \omega-1) z_{c b}-z_{c b}^2-1\right) r_\omega(\omega)\right.\nonumber\\
&&\qquad\left.-12 z_{c b}\left(\omega-\omega_{c b}\right)-\left(z_{c b}^2-1\right) \log z_{c b}\right]+V(\mu), \nonumber\\
&&\delta \hat{h}_{A_1, \alpha_s}= \frac{1}{6 z_{c b}\left(\omega-\omega_{c b}\right)}\left[4 z_{c b}\left(\omega-\omega_{c b}\right) \Omega_\omega(\omega)+2(\omega-1)\left((3 \omega+1) z_{c b}-z_{c b}^2-1\right) r_\omega(\omega)\right.\nonumber\\
&&\qquad\left.-12 z_{c b}\left(\omega-\omega_{c b}\right)-\left(z_{c b}^2-1\right) \log z_{c b}\right]+V(\mu), \nonumber\\
&&\delta \hat{h}_{A_2, \alpha_s}= \frac{1}{6 z_{c b}^2\left(\omega-\omega_{c b}\right)^2}\left[\left(2+\left(2 \omega^2-5 \omega-1\right) z_{c b}+2 \omega(2 \omega-1) z_{c b}^2+(1-\omega) z_{c b}^3\right) r_\omega(\omega)\right.\nonumber\\
&&\qquad\left.-2 z_{c b}\left(z_{c b}+1\right)\left(\omega-\omega_{c b}\right)+\left(z_{c b}^2-(4 \omega+2) z_{c b}+3+2 \omega\right) z_{c b} \log z_{c b}\right],\nonumber \\
&&\delta \hat{h}_{A_3, \alpha_s}= \frac{1}{6 z_{c b}\left(\omega-\omega_{c b}\right)^2}\left[4\left(\omega-\omega_{c b}\right)^2 z_{c b} \Omega_\omega(\omega)+\left(1+\omega-2 \omega^2+6 \omega^3 z_{c b}+z_{c b}^2\left(-1+2 z_{c b}\right)\right.\right.\nonumber\\
&&\qquad\left.-\omega z_{c b}\left(4+3 z_{c b}\right)-2(-1+\omega)\left(-1+z_{c b}+3 \omega z_{c b}-z_{c b}^2\right) \omega_{c b}\right) r_\omega(\omega) \nonumber\\
&&\qquad-2 \omega z_{c b}\left(1+6 \omega+z_{c b}\right)+\left(-10+24 \omega+2 z_{c b}\right) z_{c b} \omega_{c b} \nonumber\\
&&\qquad\left.+\left(-2+\omega+(2+4 \omega) z_{c b}-(2+3 \omega) z_{c b}^2\right) \log z_{c b}\right]+V(\mu),
\end{eqnarray}
with

\begin{eqnarray}
&&z_{c b}=\frac{m_c}{m_b}, \quad \omega_{c b}=\frac{1}{2}\left(z_{c b}+z_{c b}^{-1}\right), \quad \omega_{\pm}(\omega)=\omega \pm \sqrt{\omega^2-1}, \nonumber\\
&&r_\omega(\omega)=\frac{\log \omega_{+}(\omega)}{\sqrt{\omega^2-1}}, \nonumber\\
&&\Omega_\omega(\omega)=\frac{\omega}{2 \sqrt{\omega^2-1}}\left[2 \operatorname{Li}_2\left(1-\omega_{-}(\omega) z_{c b}\right)-2 \mathrm{Li}_2\left(1-\omega_{+}(\omega) z_{c b}\right)\right. \nonumber\\
&&\left.\quad+\operatorname{Li}_2\left(1-\omega_{+}^2(\omega)\right)-\operatorname{Li}_2\left(1-\omega_{-}^2(\omega)\right)\right]-\omega r_\omega(\omega) \log z_{c b}+1,
\end{eqnarray}
where $\mathrm{Li}_2(x)=\int_x^0 d t \log (1-t) / t$ is the dilogarithmical function. The above results are obtained at the scale $\mu_{\sqrt{b c}}=\sqrt{m_b m_c}$, namely $V\left(\mu_{\sqrt{b c}}\right)=0$. Otherwise, the scale factor is given as

\begin{eqnarray}
&&V(\mu)=-\frac{2}{3}\left(\omega r_\omega(\omega)-1\right) \log \frac{m_b m_c}{\mu^2},
\end{eqnarray}
We choose the scale  $\mu=4.2 \mathrm{GeV}$ in our calculations. The  $1 / m_{b, c}$ corrections are given as 

\begin{eqnarray}
&&\delta \hat{h}_{V, m_b} =\delta \hat{h}_{A_3, m_b}\nonumber \\
&&\qquad~~=1-2 \eta(\omega)-4(\omega-1) \hat{\chi}_2(\omega)+12 \hat{\chi}_3(\omega), \nonumber\\
&&\delta \hat{h}_{V, m_c} =1-4 \hat{\chi}_3(\omega), \nonumber\\
&&\delta \hat{h}_{A_1, m_b}=(\omega-1)\left[(\omega+1)^{-1}(1-2 \eta(\omega))-4 \hat{\chi}_2(\omega)\right]+12 \hat{\chi}_3(\omega), \nonumber\\
&&\delta \hat{h}_{A_1, m_c} =(\omega-1)(\omega+1)^{-1}-4 \hat{\chi}_3(\omega), \nonumber\\
&&\delta \hat{h}_{A_2, m_b} =\delta \hat{h}_{T_3, m_b}=0,\nonumber \\
&&\delta \hat{h}_{A_2, m_c} =-2(\omega+1)^{-1}(1+\eta(\omega))+4 \hat{\chi}_2(\omega), \nonumber\\
&&\delta \hat{h}_{A_3, m_c} =1-2(\omega+1)^{-1}(1+\eta(\omega))-4 \hat{\chi}_2(\omega)-4 \hat{\chi}_3(\omega),
\end{eqnarray}
The corrections of order $1 / m_c^2$ are included via the subleading reduced IW functions $\hat{l}_{1-6}(\omega)$ as
\begin{eqnarray}
&&\delta \hat{h}_{V, m_c^2} =\hat{\ell}_2(\omega)-\hat{\ell}_5(\omega), \nonumber\\
&&\delta \hat{h}_{A_1, m_c^2} =\hat{\ell}_2(\omega)-\frac{\omega-1}{\omega+1} \hat{\ell}_5(\omega), \nonumber\\
&&\delta \hat{h}_{A_2, m_c^2} =\hat{\ell}_3(\omega)+\hat{\ell}_6(\omega), \nonumber\\
&&\delta \hat{h}_{A_3, m_c^2} =\hat{\ell}_2(\omega)-\hat{\ell}_3(\omega)-\hat{\ell}_5(\omega)+\hat{\ell}_6(\omega).
\end{eqnarray}
for $\hat{\ell}(\omega)=\ell(\omega) / \xi(\omega)$.

\section{Fit results using the BGL parameterization with different values for $\chi_{1^+}^T(0)$ and $\tilde{\chi}_{1^-}^T(0)$}

In Tables~\ref{BGLFitLQCD1} and \ref{BGLFitLQCD2}, we present the fit results in the BGL parameterization using the ``Values II'' and ``Values III'' in Table~\ref{Inputs} as inputs.

\begin{table}[htb!]
\caption{Fit results using the BGL parameterization with experimental data. Here, the ``Values II'' in Table~\ref{Inputs} is used for $\chi_{1^+}^T(0)$ and $\tilde{\chi}_{1^-}^T(0)$.}\label{BGLFitLQCD1}
\begin{tabular}{cccc}
\hline\hline
BGL Fit &  Data +$h_{A_1}(0)=0.906$ + $\cal{B}$$(\bar{B}^0 \to D^{*+} \ell \bar{\nu}_{\ell})$ & Data + LQCD & Data + LQCD + LCSR \\
\hline
$\chi^2/\rm{d.o.f.}$ & $135.02/107$ & $166.69/144$ &  $190.66/170$  \\
$a_0^g$       & $0.022 \pm 0.009$    & $0.0240 \pm 0.0005$    & $0.0235 \pm 0.0005$    \\
$a_1^g$       & $0.045 \pm 0.344$     & $-0.057 \pm 0.016$    & $-0.093 \pm 0.020$   \\
$a_2^g$       & $-0.999_{-0.001}^{+1.999}$     & $-0.998^{+0.005}_{-0.002}$     & $0.365 \pm 0.289$      \\
$a_0^f$       & $0.0122 \pm 0.0002$    & $0.0120 \pm 0.0001$    & $0.0120 \pm 0.0001$    \\
$a_1^f$       & $0.033 \pm 0.043$    & $0.016 \pm 0.005$    & $0.008 \pm 0.005$    \\
$a_2^f$       & $-0.899_{-0.101}^{+1.899}$     & $-0.422 \pm 0.160$     & $-0.120 \pm 0.107$    \\
$a_1^{{\cal{F}}_1}$    & $0.002 \pm 0.002$    & $0.0006 \pm 0.0011$    & $0.001 \pm 0.001$    \\
$a_2^{{\cal{F}}_1}$    & $-0.016 \pm 0.031$    & $0.009 \pm 0.022$    & $-0.007 \pm 0.020$   \\
$|V_{cb}| \times 10^{-3}$ & $38.67 \pm 0.93$ & $39.68 \pm 0.56$ & $39.68 \pm 0.55$ \\
\hline\hline
\end{tabular}
\end{table}
\begin{table}[htb!]
\caption{Fit results using the BGL parameterization with experimental data. Here, the ``Values III'' in Table~\ref{Inputs} is used for $\chi_{1^+}^T(0)$ and $\tilde{\chi}_{1^-}^T(0)$. }\label{BGLFitLQCD2}
\begin{tabular}{cccc}
\hline\hline
BGL Fit &  Data +$h_{A_1}(0)=0.906$ + $\cal{B}$$(\bar{B}^0 \to D^{*+} \ell \bar{\nu}_{\ell})$ & Data + LQCD & Data + LQCD + LCSR \\
\hline
$\chi^2/\rm{d.o.f.}$ & $135.03/107$ & $196.56/144$ &  $219.41/170$  \\
$a_0^g$       & $0.025 \pm 0.010$    & $0.0257 \pm 0.0005$    & $0.0253 \pm 0.0005$    \\
$a_1^g$       & $0.044 \pm 0.370$     & $-0.060 \pm 0.018$    & $-0.097 \pm 0.021$   \\
$a_2^g$       & $-0.999^{+1.999}_{-0.001}$     & $-0.998^{+0.012}_{-0.002}$     & $0.352 \pm 0.306$      \\
$a_0^f$       & $0.0122 \pm 0.0002$    & $0.0114 \pm 0.0001$    & $0.0114 \pm 0.0001$    \\
$a_1^f$       & $0.033 \pm 0.040$    & $0.015 \pm 0.060$    & $0.007 \pm 0.051$    \\
$a_2^f$       & $-0.903 ^{+ 1.903}_{-0.097}$     & $-0.393 \pm 0.157$     & $-0.115 \pm 0.099$    \\
$a_1^{{\cal{F}}_1}$    & $0.002 \pm 0.002$    & $0.0005 \pm 0.0011$    & $0.0009 \pm 0.0011$    \\
$a_2^{{\cal{F}}_1}$    & $-0.016 \pm 0.031$    & $0.010 \pm 0.020$    & $-0.005 \pm 0.019$   \\
$|V_{cb}| \times 10^{-3}$ & $35.98 \pm 0.88$ & $39.00 \pm 0.55$ & $39.00 \pm 0.54$ \\
\hline\hline
\end{tabular}
\end{table}

\section{Correlations of fits in different parameterizations}

Here, we provide all the correlation matrices for the $|V_{cb}|$ fit in the CLN parameterization, the BGL parameterization, and the HQET~(2/1/0) parameterization.

\begin{table}[htb!]
\centering
\caption{Correlations among $\{
\rho_{D^*}^2,R_1(1),R_2(1),|V_{cb}|\}$-
$\{
\rho_{D^*}^2,R_1(1),R_2(1),|V_{cb}|\}$ in the CLN parameterization corresponding to Table \ref{CLNFit}.\label{CLNCorr}}
\begin{tabular}{ccccccc}
\\ \hline \hline
~~~ Corr.~~~ & ~~~ $\rho_{D^*}^2$~~~ & ~~~ $R_1(1)$~~~ & ~~~ $R_2(1)$~~~ & ~~~ $|V_{cb}|$~~~
\\ \hline
 
$\rho_{D^*}^2$ & $1.000$ & $-0.362$ & $-0.728$ & $-0.202$    \\
   
$R_1(1)$ & $-0.362$ & $1.000$ & $0.534$ & $-0.087$   \\
  
$R_2(1)$ & $-0.728$ & $0.534$ & $1.000$ & $-0.010$    \\
  
$|V_{cb}|$ & $-0.202$ & $-0.087$ & $-0.010$ & $1.000$    \\

 \hline \hline
\end{tabular}
\end{table}

\begin{table}[htb!]
\centering
\caption{Correlations among $\{|V_{cb}|,a_j^{F_i}\}$-$\{|V_{cb}|,a_j^{F_i}\}$ in the BGL parameterization corresponding to Table \ref{BGLFit}.\label{BGLCorr}}
\begin{tabular}{ccccccccccc}
\\ \hline \hline
~~~ Corr.~~~ & ~~~ $a_0^g$~~~ & ~~~ $a_1^g$~~~ & ~~~ $a_2^g$~~~ & ~~~ $a_0^f$~~~ & ~~~ $a_1^f$~~~ & ~~~ $a_2^f$~~~ & ~~~ $a_1^{{\cal{F}}_1}$~~~ & ~~~ $a_2^{{\cal{F}}_1}$~~~ & ~~~$|V_{cb}|$~~~ & ~~~ 
\\ \hline

$a_0^g$ & $1.000$ & $-0.445$ & $-0.084$ & $0.276$ & $-0.033$ & $-0.001$ & $0.102$ & $-0.048$ & $-0.277$   \\

$a_1^g$ & $-0.445$ & $1.000$ & $0.782$ & $0.016$ & $0.184$ & $0.174$ & $0.153$ & $-0.107$ & $-0.133$   \\

$a_2^g$ & $-0.084$ & $0.782$ & $1.000$ & $0.021$ & $0.144$ & $0.157$ & $0.096$ & $-0.042$ & $-0.114$   \\

$a_0^f$ & $0.276$ & $0.016$ & $0.021$ & $1.000$ & $-0.172$ & $-0.118$ & $-0.150$ & $0.123$ & $-0.506$   \\
     
$a_1^f$ & $-0.033$ & $0.184$ & $0.144$ & $-0.172$ & $1.000$ & $0.832$ & $0.578$ & $-0.426$ & $-0.262$   \\

$a_2^f$ & $-0.001$ & $0.174$ & $0.157$ & $-0.118$ & $0.832$ & $1.000$ & $0.369$ & $-0.264$ & $-0.142$   \\
     
$a_1^{{\cal{F}}_1}$ & $0.102$ & $0.153$ & $0.096$ & $-0.150$ & $0.578$ & $0.369$ & $1.000$ & $-0.944$ & $-0.269$   \\

$a_2^{{\cal{F}}_1}$ & $-0.048$ & $-0.107$ & $-0.042$ & $0.123$ & $-0.426$ & $-0.264$ & $-0.944$ & $1.000$ & $0.169$   \\

$|V_{cb}|$ & $-0.277$ & $-0.133$ & $-0.114$ & $-0.506$ & $-0.262$ & $-0.142$ & $-0.269$ & $0.169$ & $1.000$   \\

 \hline \hline
\end{tabular}
\end{table}

\begin{table}[h!]
\centering
\caption{Correlations among $\{|V_{cb}|,\xi^{(n)},\hat{\chi}_{2,3}^{(n)},\eta^{(n)},\hat{\ell}_{2,3,5,6}^{(0)}\}$-$\{|V_{cb}|,\xi^{(n)},\hat{\chi}_{2,3}^{(n)},\eta^{(n)}$, $\hat{\ell}_{2,3,5,6}^{(0)}\}$ in the HQET~(2/1/0) parameterization corresponding to Table~\ref{HQETFit}.\label{HQETCorr}}
\begin{tabular}{ccccccccccccccc}
\\ \hline \hline
~~~ Corr.~~~ & ~~~ $|V_{cb}|$~~~ & ~~~ $\xi^{(1)}$~~~ & ~~~ $\xi^{(2)}$~~~ & ~~~ $\hat{\chi}_2^{(0)}$~~~ & ~~~ $\hat{\chi}_2^{(1)}$~~~ & ~~~ $\hat{\chi}_3^{(1)}$~~~ & ~~~ $\eta^{(0)}$~~~ & ~~~ $\eta^{(1)}$~~~ & ~~~ $\hat{\ell}_2^{(0)}$~~~ & ~~~ $\hat{\ell}_3^{(0)}$~~~ & ~~~ $\hat{\ell}_5^{(0)}$~~~ & ~~~ $\hat{\ell}_6^{(0)}$~~~
\\ \hline
$|V_{cb}|$ & $1.000$ & $-0.263$ & $0.258$ & $0.003$ & $0.013$ & $0.002$ & $-0.017$ & $0.012$ & $-0.546$ & $0.021$ & $0.118$ & $0.073$ \\
$\xi^{(1)}$ & $-0.263$ & $1.000$ & $-0.906$ & $0.107$ & $0.007$ & $0.06$ & $0.006$ & $0.006$ & $-0.023$ & $-0.205$ & $0.329$ & $0.211$ \\
$\xi^{(2)}$ & $0.258$ & $-0.906$ & $1.000$ & $-0.051$ & $0.005$ & $-0.032$ & $0.024$ & $-0.011$ & $0.019$ & $0.092$ & $-0.152$ & $-0.072$ \\
$\hat{\chi}_2^{(0)}$ & $0.003$ & $0.107$ & $-0.051$ & $1.000$ & $-0.001$ & $0.000$ & $0.002$ & $-0.001$ & $0.005$ & $-0.31$ & $0.008$ & $0.007$  \\
$\hat{\chi}_2^{(1)}$ & $0.013$ & $0.007$ & $0.005$ & $-0.001$ & $1.000$ & $-0.001$ & $0.013$ & $-0.008$ & $0.000$ & $-0.064$ & $0.001$ & $0.03$\\
$\hat{\chi}_3^{(1)}$ & $0.002$ & $0.06$ & $-0.032$ & $0.000$ & $-0.001$ & $1.000$ & $0.001$ & $0.000$ & $0.003$ & $-0.001$ & $0.004$ & $0.004$\\
$\eta^{(0)}$ & $-0.017$ & $0.006$ & $0.024$ & $0.002$ & $0.013$ & $0.001$ & $1.000$ & $0.017$ & $-0.003$ & $0.076$ & $-0.528$ & $0.636$  \\
$\eta^{(1)}$ & $0.012$ & $0.006$ & $-0.011$ & $-0.001$ & $-0.008$ & $0.000$ & $0.017$ & $1.000$ & $-0.001$ & $-0.054$ & $-0.014$ & $0.04$  \\
$\hat{\ell}_2^{(0)}$ & $-0.546$ & $-0.023$ & $0.019$ & $0.005$ & $0.000$ & $0.003$ & $-0.003$ & $-0.001$ & $1.000$ & $-0.053$ & $0.001$ & $-0.058$ \\
$\hat{\ell}_3^{(0)}$ & $0.021$ & $-0.205$ & $0.092$ & $-0.31$ & $-0.064$ & $-0.001$ & $0.076$ & $-0.054$ & $-0.053$ & $1.000$ & $-0.199$ & $0.484$ \\
$\hat{\ell}_5^{(0)}$ & $0.118$ & $0.329$ & $-0.152$ & $0.008$ & $0.001$ & $0.004$ & $-0.528$ & $-0.014$ & $0.001$ & $-0.199$ & $1.000$ & $0.047$ \\
$\hat{\ell}_6^{(0)}$ & $0.073$ & $0.211$ & $-0.072$ & $0.007$ & $0.03$ & $0.004$ & $0.636$ & $0.04$ & $-0.058$ & $0.484$ & $0.047$ & $1.000$  \\
 \hline \hline
\end{tabular}
\end{table}

\begin{table}[htb!]
\centering
\caption{Correlations among $\{
\rho_{D^*}^2,R_1(1),R_2(1),R_0(1)\}$-
$\{
\rho_{D^*}^2,R_1(1),R_2(1),R_0(1)\}$ in the CLN parameterization corresponding to Table \ref{CLNFitLQCDLCSR} .\label{CLNLQCDLCSRCorr}}
\begin{tabular}{ccccccc}
\\ \hline \hline
~~~ Corr.~~~ & ~~~ $\rho_{D^*}^2$~~~ & ~~~ $R_1(1)$~~~ & ~~~ $R_2(1)$~~~& ~~~ $R_0(1)$~~~ 
\\ \hline

$\rho_{D^*}^2$ &  1.000 & 0.127 & -0.073 &   0.068 \\
$R_1(1)$ & 0.127 &  1.000 &  -0.203 & 0.203 \\
$R_2(1)$ & -0.073 & -0.203 &  1.000 &  -0.996 \\
$R_0(1)$ &  0.068 & 0.203 & -0.996 &   1.000 \\

 \hline \hline
\end{tabular}
\end{table}

\begin{table}[htb!]
\centering
\caption{Correlations among $\{a_j^{F_i}\}$-$\{a_j^{F_i}\}$ in the BGL parameterization corresponding to Table \ref{BGLFitLQCDLCSR} .\label{BGLLQCDLCSRCorr}}
\begin{tabular}{ccccccccccccc}
\\ \hline \hline
~~~ Corr.~~~ & ~~~ $a_0^g$~~~ & ~~~ $a_1^g$~~~ & ~~~ $a_2^g$~~~ & ~~~ $a_0^f$~~~ & ~~~ $a_1^f$~~~ & ~~~ $a_2^f$~~~ & ~~~ $a_1^{{\cal{F}}_1}$~~~ & ~~~ $a_2^{{\cal{F}}_1}$~~~&~~~ $a_0^{{\cal{F}}_2}$~~~ & ~~~ $a_1^{{\cal{F}}_2}$~~~ 
\\ \hline

$a_0^g$ & 1.000 & 0.047 & -0.257 & 0.117 & 0.011 & -0.048 & 0.037 & 0.026 & 0.244 & -0.023 \\

$a_1^g$  & 0.047 & 1.000 & -0.803 & -0.024 & 0.256 & 0.031 & 0.133 & -0.006 & 0.018 & 0.207 \\

$a_2^g$  & -0.257 & -0.803 & 1.000 & 0.003 & -0.164 & -0.077 & -0.092 & 0.007 & -0.068 & -0.132 \\

$a_0^f$  & 0.117 & -0.024 & 0.003 & 1.000 & -0.062 & 0.041 & -0.076 & 0.035 & 0.398 & -0.039 \\

$a_1^f$  & 0.011 & 0.256 & -0.164 & -0.062 & 1.000 & 0.630 & 0.567 & -0.371 & -0.026 & 0.493 \\

$a_2^f$  & -0.048 & 0.031 & -0.077 & 0.041 & 0.630 & 1.000 & 0.391 & -0.387 & 0.036 & 0.283 \\

$a_1^{{\cal{F}}_1}$  & 0.037 & 0.133 & -0.092 & -0.076 & 0.567 & 0.391 & 1.000 & -0.660 & 0.313 & 0.617 \\

$a_2^{{\cal{F}}_1}$ & 0.026 & -0.006 & 0.007 & 0.035 & -0.371 & -0.387 & -0.660 & 1.000 & -0.118 & -0.155 \\

$a_0^{{\cal{F}}_2}$  & 0.244 & 0.018 & -0.068 & 0.398 & -0.026 & 0.036 & 0.313 & -0.118 & 1.000 & -0.116 \\

$a_1^{{\cal{F}}_2}$  & -0.023 & 0.207 & -0.132 & -0.039 & 0.493 & 0.283 & 0.617 & -0.155 & -0.116 & 1.000 \\

 \hline \hline
\end{tabular}
\end{table}

\begin{table}[h!]
\centering
\caption{Correlations among $\{\xi^{(n)},\hat{\chi}_{2,3}^{(n)},\eta^{(n)},\hat{\ell}_{2,3,5,6}^{(0)}\}$-$\{\xi^{(n)},\hat{\chi}_{2,3}^{(n)},\eta^{(n)}$, $\hat{\ell}_{2,3,5,6}^{(0)}\}$ in the HQET~(2/1/0) parameterization corresponding to Table \ref{HQETFitLQCDLCSR} .\label{HQETLQCDLCSRCorr}}
\begin{tabular}{ccccccccccccccc}
\\ \hline \hline
~~~ Corr.~~~ & ~~~ $\xi^{(1)}$~~~ & ~~~ $\xi^{(2)}$~~~ & ~~~ $\hat{\chi}_2^{(0)}$~~~ & ~~~ $\hat{\chi}_2^{(1)}$~~~ & ~~~ $\hat{\chi}_3^{(1)}$~~~ & ~~~ $\eta^{(0)}$~~~ & ~~~ $\eta^{(1)}$~~~ & ~~~ $\hat{\ell}_2^{(0)}$~~~ & ~~~ $\hat{\ell}_3^{(0)}$~~~ & ~~~ $\hat{\ell}_5^{(0)}$~~~ & ~~~ $\hat{\ell}_6^{(0)}$~~~
\\ \hline
$\xi^{(1)}$  & $1.000$ & $-0.887$ & $0.074$ & $0.016$ & $0.040$ & $-0.012$ & $0.022$ & $-0.009$ & $-0.229$ & $0.205$ & $0.116$ \\
$\xi^{(2)}$  & $-0.887$ & $1.000$ & $-0.044$ & $0.006$ & $-0.025$ & $0.014$ & $-0.012$ & $-0.012$ & $0.226$ & $-0.093$ & $-0.035$ \\
$\hat{\chi}_2^{(0)}$  & $0.074$ & $-0.044$ & $1.000$ & $0.000$ & $-0.001$ & $0.000$ & $0.000$ & $0.000$ & $-0.222$ & $0.001$ & $0.000$  \\
$\hat{\chi}_2^{(1)}$  & $0.016$ & $0.006$ & $0.000$ & $1.000$ & $0.000$ & $0.003$ & $-0.002$ & $-0.002$ & $-0.011$ & $0.007$ & $0.011$\\
$\hat{\chi}_3^{(1)}$  & $0.040$ & $-0.025$ & $-0.001$ & $0.000$ & $1.000$ & $0.000$ & $0.000$ & $0.000$ & $-0.001$ & $-0.000$ & $-0.000$\\
$\eta^{(0)}$ & $-0.012$ & $0.014$ & $0.000$ & $0.003$ & $0.000$ & $1.000$ & $0.008$ & $0.000$ & $0.018$ & $-0.376$ & $0.564$  \\
$\eta^{(1)}$  & $0.022$ & $-0.012$ & $0.000$ & $-0.002$ & $0.000$ & $0.008$ & $1.000$ & $0.000$ & $-0.012$ & $0.008$ & $0.027$  \\
$\hat{\ell}_2^{(0)}$ & $-0.009$ & $-0.012$ & $0.000$ & $-0.002$ & $0.000$ & $0.000$ & $0.000$ & $1.000$ & $0.003$ & $0.084$ & $0.000$ \\
$\hat{\ell}_3^{(0)}$  & $-0.229$ & $0.226$ & $-0.222$ & $-0.011$ & $-0.001$ & $0.018$ & $-0.012$ & $0.003$ & $1.000$ & $-0.112$ & $0.070$ \\
$\hat{\ell}_5^{(0)}$ & $0.205$ & $-0.093$ & $0.001$ & $0.007$ & $-0.000$ & $-0.376$ & $0.008$ & $0.084$ & $-0.112$ & $1.000$ & $0.314$ \\
$\hat{\ell}_6^{(0)}$  & $0.116$ & $-0.035$ & $0.000$ & $0.011$ & $-0.000$ & $0.564$ & $0.027$ & $0.000$ & $0.070$ & $0.314$ & $1.000$  \\
 \hline \hline
\end{tabular}
\end{table}

\begin{table}[h!]
\centering
\caption{Correlations among $\{\xi^{(n)},\hat{\chi}_{2,3}^{(n)},\eta^{(n)},\hat{\ell}_{2,3,5,6}^{(0)}\}$-$\{\xi^{(n)},\hat{\chi}_{2,3}^{(n)},\eta^{(n)}$, $\hat{\ell}_{2,3,5,6}^{(0)}\}$ in the HQET~(2/1/0) parameterization corresponding to Table \ref{HQETFitLQCDLCSR} .}
\begin{tabular}{ccccccccccccccc}
\\ \hline \hline
 ~~~ $\xi^{(1)}$~~~ & ~~~ $\xi^{(2)}$~~~ & ~~~ $\hat{\chi}_2^{(0)}$~~~ & ~~~ $\hat{\chi}_2^{(1)}$~~~ & ~~~ $\hat{\chi}_3^{(1)}$~~~ & ~~~ $\eta^{(0)}$~~~ & ~~~ $\eta^{(1)}$~~~ & ~~~ $\hat{\ell}_2^{(0)}$~~~ & ~~~ $\hat{\ell}_3^{(0)}$~~~ & ~~~ $\hat{\ell}_5^{(0)}$~~~ & ~~~ $\hat{\ell}_6^{(0)}$~~~
\\ \hline
1.000 & -0.904 & 0.070 & 0.010 & 0.038 & -0.016 & 0.023 & -0.004 & -0.247 & 0.191 & 0.090 \\
-0.904 & 1.000 & -0.038 & 0.009 & -0.023 & 0.024 & -0.016 & -0.024 & 0.178 & -0.073 & 0.006 \\
0.070 & -0.038 & 1.000 & 0.002 & -0.001 & -0.000 & 0.001 & 0.001 & -0.258 & 0.000 & -0.001 \\
0.010 & 0.009 & 0.002 & 1.000 & 0.001 & 0.009 & -0.005 & -0.007 & -0.065 & 0.011 & 0.022 \\
0.038 & -0.023 & -0.001 & 0.001 & 1.000 & -0.000 & 0.000 & 0.001 & 0.003 & -0.000 & -0.001 \\
-0.016 & 0.024 & -0.000 & 0.009 & -0.000 & 1.000 & 0.015 & -0.008 & 0.073 & -0.383 & 0.617 \\
0.023 & -0.016 & 0.001 & -0.005 & 0.000 & 0.015 & 1.000 & 0.005 & -0.047 & 0.009 & 0.027 \\
-0.004 & -0.024 & 0.001 & -0.007 & 0.001 & -0.008 & 0.005 & 1.000 & -0.011 & 0.085 & 0.012 \\
-0.247 & 0.178 & -0.258 & -0.065 & 0.003 & 0.073 & -0.047 & -0.011 & 1.000 & -0.124 & 0.100 \\
0.191 & -0.073 & 0.000 & 0.011 & -0.000 & -0.383 & 0.009 & 0.085 & -0.124 & 1.000 & 0.291 \\
0.090 & 0.006 & -0.001 & 0.022 & -0.001 & 0.617 & 0.027 & 0.012 & 0.100 & 0.291 & 1.000 \\
 \hline \hline
\end{tabular}
\end{table}

\begin{table}[htb!]
\centering
\caption{Correlations among $\{|V_{cb}|,a_j^{F_i}\}$-$\{|V_{cb}|,a_j^{F_i}\}$ in the BGL parameterization corresponding to Table\ref{BGLFitLQCD1}.\label{BGLCorrII}}
\begin{tabular}{ccccccccccc}
\\ \hline \hline
~~~ Corr.~~~ & ~~~ $a_0^g$~~~ & ~~~ $a_1^g$~~~ & ~~~ $a_2^g$~~~ & ~~~ $a_0^f$~~~ & ~~~ $a_1^f$~~~ & ~~~ $a_2^f$~~~ & ~~~ $a_1^{{\cal{F}}_1}$~~~ & ~~~ $a_2^{{\cal{F}}_1}$~~~ & ~~~$|V_{cb}|$~~~ & ~~~ 
\\ \hline

$a_0^g$ & $1.000$ & $-0.434$ & $0.050$ & $0.284$ & $0.006$ & $-0.032$ & $0.136$ & $-0.072$ & $-0.293$   \\

$a_1^g$ & $-0.434$ & $1.000$ & $-0.767$ & $0.044$ & $0.186$ & $-0.172$ & $0.150$ & $-0.104$ & $-0.143$   \\

$a_2^g$ & $0.050$ & $-0.767$ & $1.000$ & $-0.049$ & $-0.145$ & $0.158$ & $-0.095$ & $0.039$ & $0.128$   \\

$a_0^f$ & $0.284$ & $0.044$ & $-0.049$ & $1.000$ & $-0.011$ & $-0.016$ & $-0.045$ & $0.042$ & $-0.549$   \\
     
$a_1^f$ & $0.006$ & $0.186$ & $-0.145$ & $-0.011$ & $1.000$ & $-0.819$ & $0.544$ & $-0.389$ & $-0.332$   \\

$a_2^f$ & $-0.032$ & $-0.172$ & $0.158$ & $-0.016$ & $-0.819$ & $1.000$ & $-0.324$ & $0.220$ & $0.195$   \\
     
$a_1^{{\cal{F}}_1}$ & $0.136$ & $0.150$ & $-0.095$ & $-0.045$ & $0.544$ & $-0.324$ & $1.000$ & $-0.944$ & $-0.309$   \\

$a_2^{{\cal{F}}_1}$ & $-0.072$ & $-0.104$ & $0.039$ & $0.042$ & $-0.389$ & $0.220$ & $-0.944$ & $1.000$ & $0.198$   \\

$|V_{cb}|$ & $-0.293$ & $-0.143$ & $0.128$ & $-0.549$ & $-0.332$ & $0.195$ & $-0.309$ & $0.198$ & $1.000$   \\

 \hline \hline
\end{tabular}
\end{table}

\begin{table}[htb!]
\centering
\caption{Correlations among $\{|V_{cb}|,a_j^{F_i}\}$-$\{|V_{cb}|,a_j^{F_i}\}$ in the BGL parameterization corresponding to Table~\ref{BGLFitLQCD2}.\label{BGLCorrIII}}
\begin{tabular}{ccccccccccc}
\\ \hline \hline
~~~ Corr.~~~ & ~~~ $a_0^g$~~~ & ~~~ $a_1^g$~~~ & ~~~ $a_2^g$~~~ & ~~~ $a_0^f$~~~ & ~~~ $a_1^f$~~~ & ~~~ $a_2^f$~~~ & ~~~ $a_1^{{\cal{F}}_1}$~~~ & ~~~ $a_2^{{\cal{F}}_1}$~~~ & ~~~$|V_{cb}|$~~~ & ~~~ 
\\ \hline

$a_0^g$ & $1.000$ & $-0.427$ & $0.046$ & $0.287$ & $0.004$ & $-0.033$ & $0.135$ & $-0.071$ & $-0.294$   \\

$a_1^g$ & $-0.427$ & $1.000$ & $-0.768$ & $0.044$ & $0.187$ & $-0.172$ & $0.151$ & $-0.105$ & $-0.143$   \\

$a_2^g$ & $0.046$ & $-0.768$ & $1.000$ & $-0.049$ & $-0.146$ & $0.158$ & $-0.095$ & $0.040$ & $0.128$   \\

$a_0^f$ & $0.287$ & $0.044$ & $-0.049$ & $1.000$ & $-0.012$ & $-0.016$ & $-0.047$ & $0.043$ & $-0.550$   \\
     
$a_1^f$ & $0.004$ & $0.187$ & $-0.146$ & $-0.012$ & $1.000$ & $-0.817$ & $0.546$ & $-0.391$ & $-0.329$   \\

$a_2^f$ & $-0.033$ & $-0.172$ & $0.158$ & $-0.016$ & $-0.817$ & $1.000$ & $-0.326$ & $0.223$ & $0.192$   \\
     
$a_1^{{\cal{F}}_1}$ & $0.135$ & $0.151$ & $-0.096$ & $-0.047$ & $0.546$ & $-0.326$ & $1.000$ & $-0.943$ & $-0.305$   \\

$a_2^{{\cal{F}}_1}$ & $-0.071$ & $-0.105$ & $0.040$ & $0.043$ & $-0.391$ & $0.223$ & $-0.943$ & $1.000$ & $0.195$   \\

$|V_{cb}|$ & $-0.294$ & $-0.143$ & $0.128$ & $-0.550$ & $-0.329$ & $0.192$ & $-0.305$ & $0.195$ & $1.000$   \\

 \hline \hline
\end{tabular}
\end{table}

\bibliography{Vcb}

\end{document}